\documentclass[useAMS,usenatbib]{mn2e}

\usepackage{lscape}
\usepackage{graphicx}
\usepackage{amsmath}
\usepackage{amsfonts}

\newcommand\torus {\textsc{torus} }
\newcommand\vturb {$v_{\textrm{turb}}$}
\newcommand\mvturb {v_{\textrm{turb}}}
\newcommand\nnh {N$_{2}$H$^{+}$}
\newcommand\nnhs {N$_{2}$H$^{+}$ }

\newcommand\lows {$^{13}$CO }
\newcommand\highs {C$^{18}$O }
\newcommand\low {$^{13}$CO}
\newcommand\high {C$^{18}$O}
\newcommand\rcrit {$r_{\textrm{crit}}$ }
\newcommand\mrcrit {r_{\textrm{crit}}}
\newcommand\hmax {$h_{\textrm{max}}$ }

\newcommand\tff {t$_{\textrm{ff}}$ }

\usepackage{multirow}
\usepackage{subfigure}

\title[Three-dimensional molecular line transfer]{Three-dimensional molecular line transfer: A simulated star-forming region} \author[D. Rundle et al.]{David Rundle\thanks{E-mail: drundle@astro.ex.ac.uk}, Tim J. Harries, David M. Acreman and Matthew R. Bate\\
School of Physics,
  University of Exeter, Stocker Road, Exeter, EX4 4QL.\\
 }
 
\begin{document}

\date{Accepted ?. Received ?; in
  original form ?}

\pagerange{\pageref{firstpage}--\pageref{lastpage}} \pubyear{2010}

\maketitle

\label{firstpage}

\begin{abstract} 
We present the first non-LTE, co-moving frame molecular line calculations of a star-forming cluster simulated using smoothed particle hydrodynamics (SPH), from which we derive high-resolution synthetic observations. We have resampled a particle representation onto an adaptive mesh and self-consistently solved the equations of statistical equilibrium in the co-moving frame, using \textsc{torus}, a three-dimensional adaptive mesh refined (AMR) radiative transfer (RT) code.
We verified the applicability of the code to the conditions of the SPH simulation by testing its output against other codes. We find that the level populations obtained for optically thick and thin scenarios closely match the ensemble average of the other codes.
We have used the code to obtain non-LTE level populations of multiple molecular species throughout the cluster and have created three-dimensional velocity-resolved spatial maps of the emergent intensity.
Line profiles of cores traced by \nnh~(1-0) are compared to probes of low density gas, \low~(1-0) and \high~(1-0), surrounding the cores along the line of sight. The relative differences of the line-centre velocities are shown to be small compared to the velocity dispersion, matching recent observations. We conclude that one cannot reject competitive accretion as a viable theory of star formation based on observed velocity profiles.
\end{abstract}

\begin{keywords}
  radiative transfer -- methods: numerical -- stars: formation -- ISM: molecules -- ISM: kinematics and dynamics
\end{keywords}

\section{Introduction}
\label{Intro}
Large-scale hydrodynamic collapse calculations of turbulent clouds (e.g. \citealt*{BaBoBromm03,KruKleMcK07a}) demonstrate the fragmentation of molecular clouds into clusters which in turn typically fragment into high-density cores and subsequently fragment into protostars. These simulations predict the statistical properties of the distribution of young stellar objects such as the initial mass function, the binary fraction and the star formation rate (e.g. \citealt{Zin84,BoBaVi03,BaBoBromm03,Bate09a}). It is expected that the fate of these objects is strongly affected by their evolution history, specifically the competitive accretion of gas as they move through their environs. Star formation in isolation \citep*{AdamsLadaShu87} necessarily neglects the dynamic interactions between these objects, which can cause them to be ejected at great velocities from their nascent gas-rich core, truncating their accretion \citep*{Klessen98,BaBoBromm02b} and it will be possible, using optically thin molecular tracers, to observe any motion of the object relative to the enveloping gas \citep*{Walshetal04}.

Comparisons of the collapse calculations with observations have generally relied on statistical tests of protostellar properties such as the IMF and binary fractions (\citealt{Bate09b}, but see also \citealt{Kurosawaetal04}). Although undoubtedly useful, such comparisons neglect the information encoded in the geometrical and dynamical properties of the star-forming gas and considering the wealth of observational data available (e.g. \citealt{NarHeyBru08,BucCurRob10}) these tests should provide a further, robust appraisal of the models.
In order to make comparisons between theory and observation, given the intractability of the inverse problem, it is necessary to transform the temperature and density characteristics of a simulated cloud into a synthetic intensity map comparable with those obtained by observation. That being so, this paper aims to improve upon the analysis of the relative motions of cores to their nascent envelopes in the cluster models of \citet{BaBoBromm03} by \citet{Ayliffeetal07}. Both \cite{Ayliffeetal07} and \cite{Walshetal04} look for evidence of supersonic ballistic motions of cores relative to their inchoate envelopes. Assuming that the velocities of the core and the envelope are in the line of sight it is possible to determine the difference between them.  By performing a non-LTE  radiative transfer calculation of the molecular level populations within the cluster using \textsc{torus} (see \citealt{Harries00, Harries04, Symington05a} for details), we seek to improve upon the comparison between simulation and observation carried out by \cite{Ayliffeetal07} and reconcile the differences between observation and theory.

In order to understand the chemical and physical conditions in these structures it is necessary to understand the role that molecules play in radiation transfer in these cool, dense regions. For over 40 years (e.g. \citealt{Dieter64, WilsonPenzias70}) molecular line radiation has regularly been used to trace key physical parameters such as density, temperature and velocity. Today's (sub-)millimetre observatories enable the observation of some of the coolest, most chemically rich regions of the galaxy. Furthermore, the Atacama Large Millimeter Array (ALMA) promises unprecedented resolution of these objects therefore it is vital that the interpretation of the data is sound, necessitating the use of non-symmetric models incorporating complex physics. This was first recognized by \citet{Bernes79} and today many line radiative transfer codes exist (e.g. \citealt{Juvela97, Wiesemeyer97, Rawlings&Yates01}). Until relatively recently, only spherically or axially symmetric core collapse simulations of line transfer had been performed (e.g. \citealt{Tsamis08,Pavetal08})
however, due to recent advances in computational power, it is now possible to compute three-dimensional line transfer for these complex density structures in star-forming regions. Here we use the \torus code, since the use of adaptive mesh refinement (AMR) enables us to treat the huge dynamic range in density and linear scale associated with the SPH cluster calculations.

In section \ref{Implementation} we present our implementation of a non-LTE molecular line radiative transfer module for \textsc{torus}. In section \ref{benchmarking} we demonstrate agreement with other codes in some of the benchmark problems set out in \citet{vZ02} and verify the accuracy of our raytracing routine for generating datacubes. In section \ref{sphtogrid}, we present an efficient method for mapping irregular SPH data on to an AMR grid and in section \ref{sph} we present the results of its application to the hydrodynamic clustered star-formation simulation of \cite{BaBoBromm03}. Finally, we discuss the results of our analysis of the cluster simulation and conclude on our findings and on the importance of radiative transfer (RT) analysis of star formation as a whole.

\section{Method}
\label{Implementation}
In order to accurately determine the local radiation field due to the presence of molecules, our radiative transfer code adopts the accelerated Monte-Carlo (AMC) method described in \citet{H&vdTak00}. The greatest benefit of AMC methods is the ability to solve the equations of radiative transfer in regions of exceptionally high optical depth ($\tau > 100$). In particular, \citeauthor*{H&vdTak00}'s method uses rays, or long characteristics, to sample the external radiation field as opposed to tracking photon packets, c.f. \citet{Lucy99}, which can become trapped leading to poor convergence or sometimes failure to converge at all. Using this method, each grid cell is sampled individually meaning that no cell is undersampled. Furthermore, the separation of local and external contributions to the radiation field in a cell facilitates convergence even in optically thick regions. This is analogous to the operator-splitting step in Accelerated Lambda Iteration (ALI) methods (see \citealt{RybHum91} for details).

\subsection{Determining non-LTE level populations}
In the scheme, there are two stages; the first stage samples the intensity of radiation incident upon each cell systematically using a small set of rays (we use 100) each possessing a unique origin within the cell, direction ($\mathbf{d}$) and a frequency ($\nu$). The frequency is sampled from a uniform distribution with a width of 4.3 normalized turbulent line widths, $v_{\textrm{turb}}\nu_{0}/c$, centred on the rest frequency for the transition, $\nu_{0}$. Beyond 2.15 linewidths, the line profile function, $\phi({\nu})$, drops to less than 1 per cent of its peak value. The distribution from which the frequency is picked is uniform to ensure good sampling in the optically-thin line-wings. In the first stage, the direction and frequency of the set of rays do not vary from iteration to iteration, nullifying the random fluctuations in coverage of the radiation field and consequently converges rapidly towards a self-consistent solution. However, because the radiation field is poorly sampled both spatially and in frequency, the solution will be strongly dependent upon the choice of initial random seed. 

The second stage of the scheme reduces these systematic and random errors by doubling the number of rays used to sample the radiation field per cell per transition with each successive iteration. The noise in a solution from a Monte-Carlo simulation diminishes proportional to $\sqrt{N}$ asymptotically, where $N$ is the number of rays used to sample the field, so it is quite possible to have to sample a cell with in excess of 10$^{5}$ rays in order to reach an acceptable level of convergence. Moreover, by allowing the properties of each ray to vary from iteration to iteration any source of systematic error will also be reduced. 

\citet{Juvela97} describes a number of schemes for reducing the noise and accelerating the convergence of Monte-Carlo methods. We have used low-discrepancy sequences to optimally cover the direction/frequency parameter space and have implemented Ng acceleration \citep{Ng74} with weights from \cite*{OAB86} to improve the convergence properties of our code without loss of generality. We have not yet implemented any kind of spatial weighting to reduce the variance between iterations for the code and as such it is possible that the contribution of a small bright grid cell to another cell at some distance will often be missed when the number of rays used per cell to sample the field is low. However, in the scenarios considered in this work, it has not been necessary to implement such a strategy.

\subsubsection{Determining the external radiation field} 
Common to both stages is the calculation of the mean intensity, $\bar{J}_{\nu}$, experienced by each cell in the grid. This is determined by averaging the contribution of samples of the specific intensity incident at a point in the cell over $4\pi$ steradians. The specific intensity, $I_{\nu}$, upon the point of origin of the ray then follows from integrating the radiative transfer equation, 
\begin{equation} \label{eqn:didtau} \frac{dI_{\nu}}{d\tau_{\nu}} = -I_{\nu} + S_{\nu},
\end{equation}
where 
\begin{equation} \label{eqn:atotau} d\tau_{\nu} = \alpha_{\nu}ds 
\end{equation}
and
\begin{equation} \label{eqn:Snu} S_{\nu} = \frac{j_{\nu}}{\alpha_{\nu}}
\end{equation}
in a piecewise manner along the ray direction towards the edge of the grid, the contribution of each subsequent cell being attenuated by the frequency specific optical depth, $\tau_{\nu}$, of all intervening cells. The source function, $S_{\nu}$, is defined as the ratio of the emission coefficient, $j_{\nu}$ (erg s$^{-1}$ cm$^{-3}$ Hz$^{-1}$ sr$^{-1}$), to the absorption coefficient, $\alpha_{\nu}$ (cm$^{-1}$). We assume the blackbody radiation from the cosmic microwave background with characteristic temperature $T_{\textrm{CMB}}=2.73$~K, $B(T_{\textrm{CMB}}$, $\nu_0$) as an external boundary condition, where $B$ is the Planck function.

All physical quantities within the cell are assumed to be constant except for the systematic velocity vector, \textbf{v(x)}, which enters equations (\ref{eqn:atotau},\ref{eqn:Snu}) through the line function $\phi(\nu)$ in both the emission and absorption coefficients (in the absence of dust):
\begin{align} 
\label{eqn:jnu} j_{\nu} &= \frac{h\nu_{0}}{4\pi}(n_{u}A_{ul}) \phi_{\nu}\\
\label{eqn:alphanu} \alpha_{\nu} &= \frac{h\nu_{0}}{4\pi}(n_{l}B_{lu} - n_{u}B_{ul}) \phi_{\nu}.
\end{align}
${A_{ul}}$, $B_{ul}$ and $B_{lu}$ are the Einstein coefficients of spontaneous emission, stimulated emission and stimulated absorption associated with the transition from level $u$ to level $l$ respectively; similarly $n_{u}$ and $n_{l}$ are the respective number densities of molecules populating those levels. Absorption due to dust can be parameterised by $\kappa_{\nu}(T)$ which may also vary as a function of position. Reprocessed dust emission at the same frequency is determined by the Planck function at the dust temperature, $T_{\textrm{dust}}$ which is assumed to be equal to the gas temperature unless otherwise stated.

The line profile function $\phi(\nu)$, centred on frequency $\nu_{0}$, is characterized by the line broadening parameter \vturb:
\begin{equation} \label{eqn:phi} \phi_{\nu} = \frac{c}{\mvturb\nu_{0}\sqrt{\pi}} \textrm{exp}\left(- {\frac{\Delta v^2}{\mvturb^2}} \right),
\end{equation}
\begin{equation} \label{eqn:microturb} \mvturb = \sqrt{v^2_{\textrm{T}} + v^2_{\textrm{NT}}},
\end{equation}
recalling that the thermal broadening $v^2_{\textrm{T}} = \frac{kT}{m}$ where $m$ is the molecular mass and where $v^2_{\textrm{NT}}$ is used to parameterize the unresolved non-thermal component of the gas velocity. $\Delta v$ is the equivalent velocity required to Doppler-shift $\nu_{0}$ to $\nu$. However the presence of a systematic velocity field introduces anisotropy by its inclusion in $\Delta v$, thus: 
\begin{equation} \label{eqn:delphav} \Delta v = c \frac{\nu - \nu_{0}}{\nu_0} + \mathbf{v} \cdot \mathbf{d}.
\end{equation}

\torus accepts an analytical function for evaluation of $\mathbf{v(x)}$. However, in the absence of one, or if the velocity is computationally expensive to calculate, it is possible to quadratically interpolate over values stored at the grid cell corners. The local source function can vary strongly with velocity gradient across a cell when the velocity gradient is large compared to the local magnitude of the turbulent velocity field. Variations in $\phi(\nu)$ are determined by subdividing the integration into smaller steps. To ensure good resolution of the velocity structure and consequently the differential optical depth within a cell, we use the condition,
\begin{equation}
n_{\textrm{split}} = \max \left(2, \lfloor 5 \frac{|\mathbf{v}_e - \mathbf{v}_s|}{v_{\textrm{turb}}}\rfloor \right)
\end{equation}
where $\mathbf{v}_e$ and $\mathbf{v}_s$ are the velocities at the cell boundaries where the ray intersects with the cell.

Equation \ref{eqn:didtau} is integrated along each ray from cell boundary to cell boundary to the grid edge for each transition being considered to build up a representation of the radiation field external to the cell (as well as quantifying the extent to which it has been attenuated by intervening material), which when added to contributions from within the cell approximates the mean radiation field.

\subsubsection{Determining $\bar{J}$ and solving the equations of statistical equilibrium}	
The mean radiation field, $\bar{J}_{\nu}$, for the cell is determined by averaging the contribution from each ray, weighted by its line profile function. $\bar{J}_{\nu}$ is split into an external component whose value is fixed and an internal component that varies according to the relative ratios of the level populations which affect $S_\nu$ so that
\begin{equation} 
\label{eqn:bigjnu} 
\bar{J}_\nu = J_\nu^{\textrm{ext}} + J_\nu^{\textrm{int}} = \frac{\sum_i I_{\nu}^{i}e^{-\tau_i} \phi_{\nu}}{\sum_i\phi_{\nu}}  +  \frac{\sum_i S_{\nu}[1 - e^{-\tau_i}] \phi_{\nu}}{\sum_i\phi_{\nu}}.
\end{equation}

In order to obtain the instantaneous relative fractional level populations, $\mathbf{n}_{i}$, for a cell it is necessary to solve a system of equations describing the statistical equilibrium that is to be attained, the so-called equations of detailed balance, viz.,	
\begin{align} 
\label{eqn:balance} n_l \left[\sum_{k<l} A_{lk} + \sum_{k\not=l}(B_{lk}J_{\nu} + C_{lk})\right] = \nonumber \\
\sum_{k>l}n_kA_{kl} + \sum_{k\not=l}n_k(B_{kl}J_\nu + C_{kl})
\end{align}
where $C_{lk}$ are the total rate coefficients of collision from level $l$ to $k$ for the molecule and a particular collision partner (e.g. H$_2$ or $e^-$) at a specific temperature.

The results of non-LTE calculations are known to be highly sensitive to the quality of these coefficients so good estimates are essential. \torus uses molecular data ($\nu_{0},A_{ul},B_{ul},B_{lu},C_{lu}(T)$) from the LAMDA database \citep{Schoier05}, a repository of molecular and atomic data for many common species.

For each cell, the system of equations (\ref{eqn:bigjnu}, \ref{eqn:balance}) are solved iteratively to obtain a self-consistent solution vector of relative level populations. \torus uses a user-defined parameter to specify the point at which successive iterations are determined to have converged. By default, the criterion is set so that the root-mean-square (RMS) error between the two most recent iterations over all but the two uppermost levels is less than $10^{-10}$, i.e. 
\begin{equation}
\label{eqn:convergence}
\sqrt{\sum_{i}\left({\frac{n_{\textrm{new}}^i - n_{\textrm{old}}^i}{n_{\textrm{new}}^i}}\right)^2} < 10^{-10}.
\end{equation}
Typically, the uppermost levels are chosen so that their population has no significant impact on the quality of the solution of lower levels. Due to their low population, the uppermost levels are often subject to a high degree of noise and their convergence is not representative of the convergence of the other levels. This criterion ensures every important level is converged to at least $10^{-10}$ and many will be converged to a much higher degree.

\subsubsection{Convergence and acceleration}
Once each cell has an updated set of level populations, they are compared with those from the previous iteration. The grid is considered to be converged when the RMS error over all cells is less than a user-specified tolerance, typically 1 per cent, for all levels.

Once the grid has converged using fixed sets of rays, the algorithm enters the second stage using increasingly large sets of rays to sample the mean radiation field. Each iteration may potentially take a long time to complete; consequently, at the end of each iteration the intermediate grid is saved along with diagnostic information that can be useful to determine if there are pockets of slower convergence in an otherwise converged grid.

As with many problems that are solved numerically by repeated function evaluation it is often possible to obtain a more accurate solution more rapidly by employing vector sequence acceleration. \torus uses a vector acceleration technique developed by \citet{Ng74} to extrapolate an updated set of relative level populations from the previous four iterations of $\mathbf{n}_i$. \cite{OAB86} comment that weighting each level with $W_{i} = {J_{ul}}^{-1}$ ensures regions with small $J$ are adequately represented. A further advantage of this method is that because the vector is taken from a space where $\sum{\mathbf{n}_i} = 1$, the relative level populations will still be normalized. In a complex iterative scheme such as this one where function evaluation can be expensive, we find that these acceleration techniques, used appropriately, can not only reduce the time required to find an accurate solution but may, in cases of extremely high optical depth, be used to determine a solution that might not be possible to obtain by fixed iteration at all. 
\subsection{Datacube generation}
\label{raytracing}
As well as determining the non-LTE level populations of a simulation, \torus can create three-dimensional velocity-resolved spatial maps of the emergent intensity (hereinafter described as \textit{datacubes}).
Datacubes of any size may be created, limited only by the amount of memory and the spatial resolution of the original AMR grid. Additionally, \torus stores extra information such as the optical depth reached in a particular datacube element and the column density. \torus generates datacubes in a similar way to that in which it finds a non-LTE solution for the level populations within a grid. The incident radiation at a point is determined by querying the source function along the path of the ray, except that when creating datacubes the point lies outside the grid, on an image plane centred on the hypothetical observer's position and that now the frequency of the observation is chosen by the observer. Typically, the datacube is comprised of many velocity channels which the observer may scan through in order to create images of intensity at each corresponding frequency. \citet{Douglasetal10} have extended this code to render images in solid angle (as opposed to rectilinear coordinates) in the near field where the observer is close to or actually inside the object, that is, observing the spiral structure of a galaxy in H\textsc{i} from within the galaxy itself.

The plane of projection is defined by the vector passing through the point of observation and a point inside the grid. Orthonormal basis vectors for the image are chosen such that one is perpendicular to $\hat{z}$ (in the case where the observer is looking along $\hat{z}$, the bases are chosen as $\pm \hat{x}$ and $\pm \hat{y}$). The image is then subdivided into $N^2$ square pixels of equal area so that the intensity per pixel may be stored in the datacube structure.

More often than not, the resolution of the image will be less than that of the grid being imaged, i.e. the width of a pixel will be greater than that of the smallest subcells contained within the grid. This necessarily means that some information about the structure will be lost. This is especially likely to occur if the intensity in a pixel is assumed to be that calculated by the pencil-beam emanating from the centre of the pixel integrated over the area of the pixel. \torus utilizes sub-pixel sampling to ameliorate this effect, by keeping a running average of the intensity sampled over different paraxial rays bound within the pixel until the standard error is less than some user-specified amount (we use 1 per cent) or until the number of rays exceeds a pre-specified maximum.

The origin of the ray inside the pixel is chosen using a Sobol quasi-random number generator that is reset for each new pixel. The advantages over pseudo-randomly generated origins are two-fold; intuitively, the first ray always originates at the pixel centre, and, the origins are created so that they avoid each other as best possible within the confines of the pixel thereby avoiding clumpiness (mathematically, a property known as low discrepancy). This provides a truer representation of the range of intensities in the region being sampled. Furthermore, because these low discrepancy sequences are capable of being extended in a pointwise fashion until some criterion is met, they are considerably more flexible than grid-based subsampling techniques that require an exponential increase of samples. Moreover, in geometries where the density can be evaluated analytically or is stored on the corners of each cell in addition to the cell centres, \torus uses density subsampling.

In the iterative level population solver, a ray is subdivided within a cell to ensure good resolution of the velocity field. Analogously, using this technique it is possible to use a more accurate value for the density at a point along the ray than that of the cell centre itself producing both smoother and more accurate images at lower grid resolutions. Furthermore, whilst errors in determining the intensity from very optically thick regions have little effect upon the level populations in their neighbourhood, they can strongly affect the appearance of an object because they affect the shape of the boundary that can be probed by radiation of a particular wavelength. 

\section{Benchmarking}
\label{benchmarking}

In order to verify the accuracy of a code, it is necessary to compare its output against either analytical solutions or the results of other codes. We tested \torus against a one-dimensional collapsing cloud model to benchmark the iterative solver used to ascertain the non-LTE level population solution along the radius of the cloud. In order to test the datacube generation, we chose a two-dimensional, very optically thick circumstellar disc model in LTE. We compared this against results obtained by \textsc{mcfost} \citep{Pinteetal06, PinHarMin09}.

\subsection{A collapsing cloud in HCO$^{+}$}
\label{molebench}

This problem was first presented as a robust test of a molecular line radiative transfer code's ability to deal with as many astrophysical phenomena as possible at a 1999 workshop on `Radiative Transfer in Molecular Lines' held in Leiden. It has since become a standard test of a molecular line transfer code's ability to reproduce level populations in a complex physical situation. Although the spherically symmetric model is quite straightforward to implement, the velocity and temperature gradients, variable turbulent line widths and multiple levels provide a stern test of a code's accuracy, especially at higher optical depths. It is noted that whilst no analytic solution exists for this benchmark, multiple codes have reached a broad consensus (within 20 per cent, though often much better depending on the fractional abundance of HCO$^{+}$ and the specific level being verified) as to what the level populations should be along the radius of the cloud. The results of this test were published in \citet{vZ02}.

\begin{figure}
  \includegraphics[width=0.475 \textwidth]{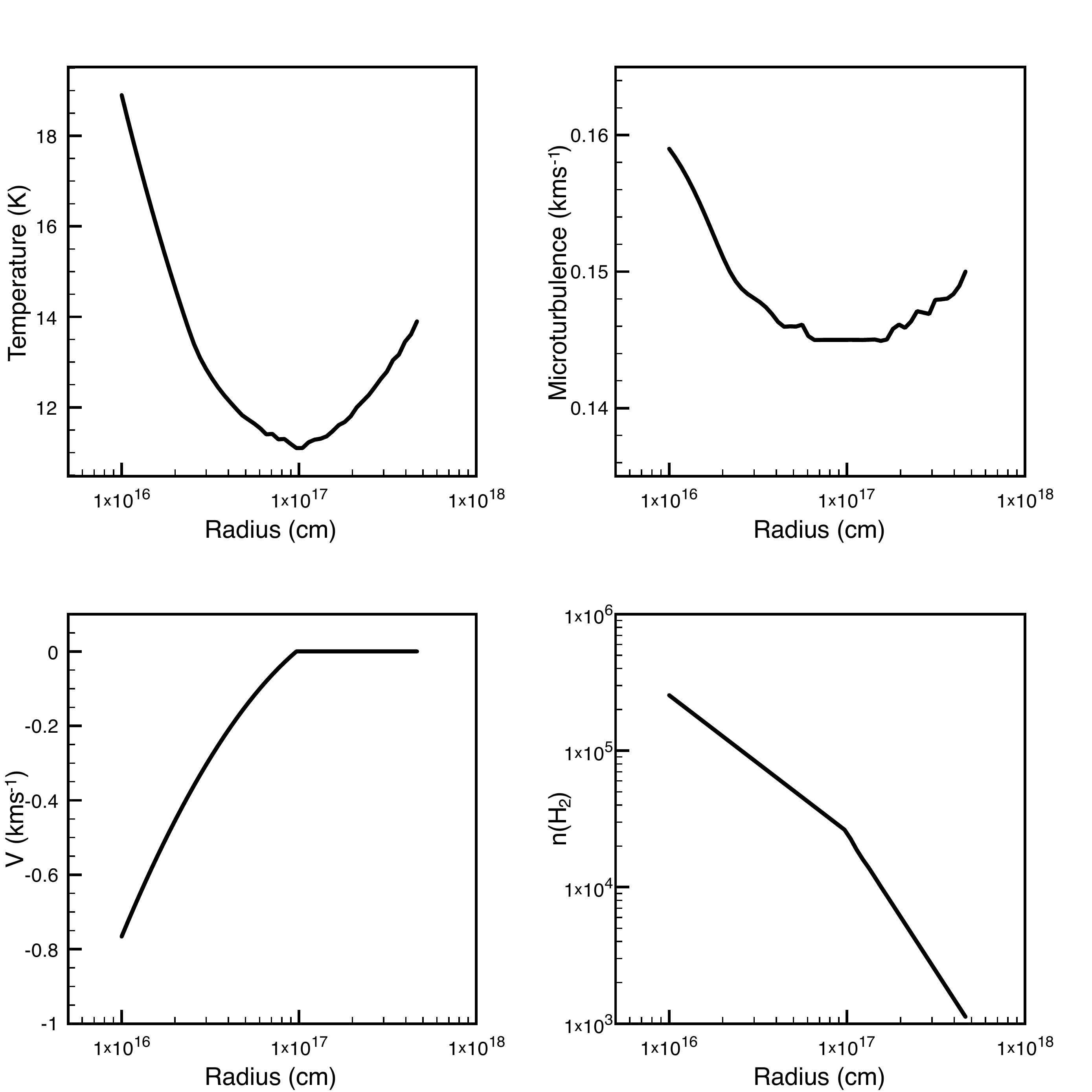}
  \caption{Parameters as a function of radius along the cloud. The stark change in $n$(H$_2)$, temperature and velocity at $10^{17}$ cm is evidence of the collapse front having reached this radius.}
  \label{fig:molebenchparams}
\end{figure}

The problem is based on a model by \citet{Rawlingsetal92} to analyse HCO$^{+}$ data for an infalling envelope around a protostar. The model is similar to the inside-out collapse model theorized by \citet{Shu77}. In this problem, the collapse is signified by the negative radial velocity in the cloud interior. Because the collapse can only propogate outward at the sound speed in the cloud, the outer shell of the cloud is static (see Figure~\ref{fig:molebenchparams}). The density profile follows a piecewise power-law $n(r) = n_{0}(r/r_{0})^{-m}$ where $r_{0} = 10^{17}$~cm and $m = 1.5$ for $r < r_{0}$ and $m = 2$ beyond $r_{0}$. All other parameters are specified at 50 logarithmically spaced points between 10$^{16}$ cm and 4.6$\times10^{17}$~cm except the relative abundance of HCO$^{+}$ to H$_{2}$ which is constant at either [HCO$^{+}$] = 10$^{-9}$ (model (a)) or [HCO$^{+}$] = 10$^{-8}$ (model (b)).

\torus uses an AMR grid which is refined according to a series of user-specified rules. In this problem, where the input parameters are tabulated, this grid is split so that no cell contains more than one datum point as defined in the source model. In each octal, the salient parameters, temperature, $n$(H$_2$), systematic velocity field and local turbulent velocity are assigned assuming conditions at the centre of the cell apply throughout its extent. In this problem both thermal and turbulent velocities are combined. Where the cell centre does not coincide with a datum point the parameters are logarithmically interpolated where an analytical value is not available.

To maintain consistency with other results we used the molecular data supplied with the problem set; a database of Einstein coefficients and collisional rate coefficients of HCO$^{+}$ with H$_{2}$ in $J=0$ (\citealt{Monteiro85,Green75}). For radii inside and outside the region covered by the data file we assumed a vacuum at $T_{\textrm{CMB}}$.

\begin{figure*}
\includegraphics[width=0.55 \textwidth]{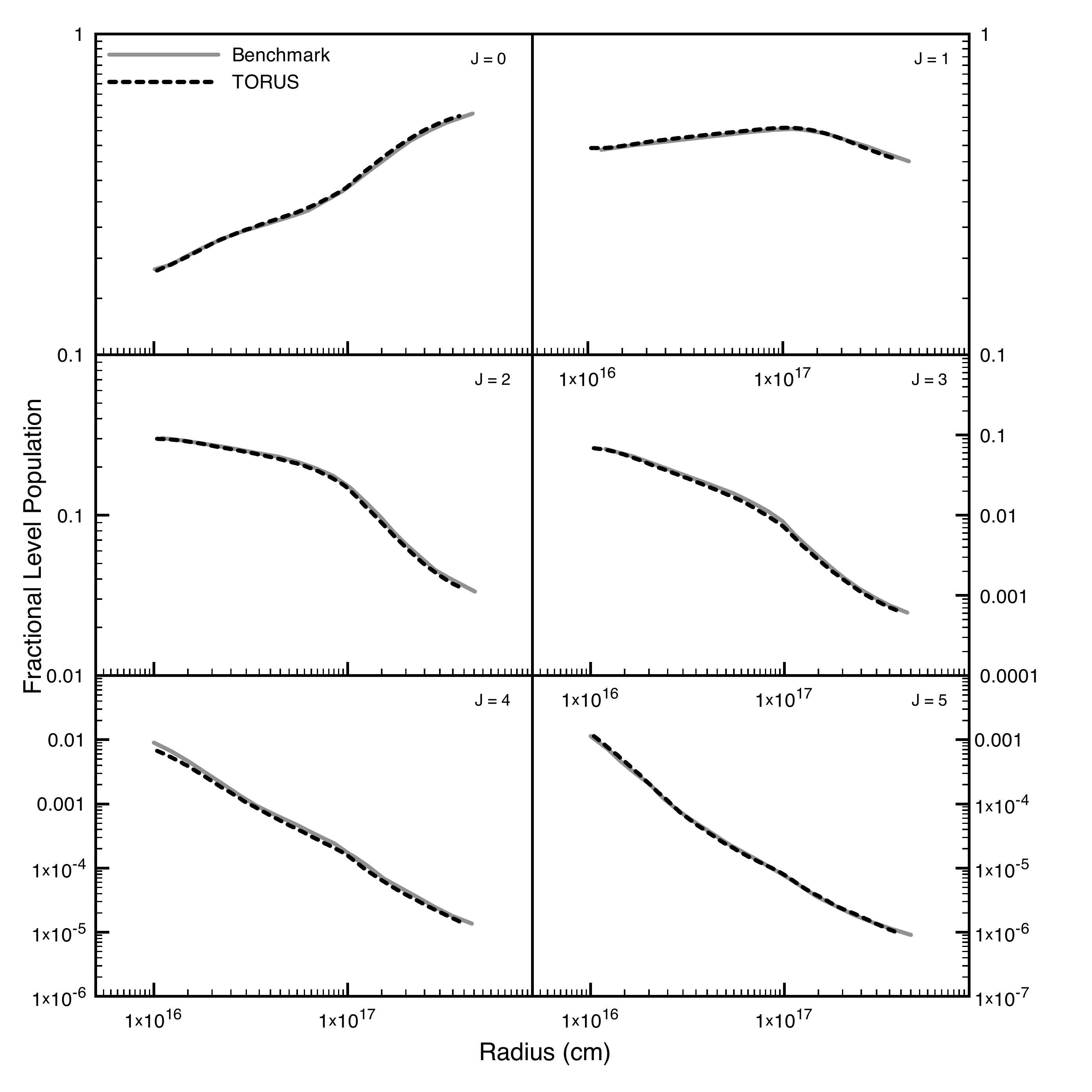}
\caption{
Relative level populations for $J=0$ to $J=5$ for the optically thin benchmark case ($\tau_{J=2-1} \approx 6.5$) where [HCO$^{+}$] = 10$^{-9}$.
}
\label{fig:fracpops1e-9}
\end{figure*}

\begin{figure*}
\includegraphics[width = 0.55 \textwidth]{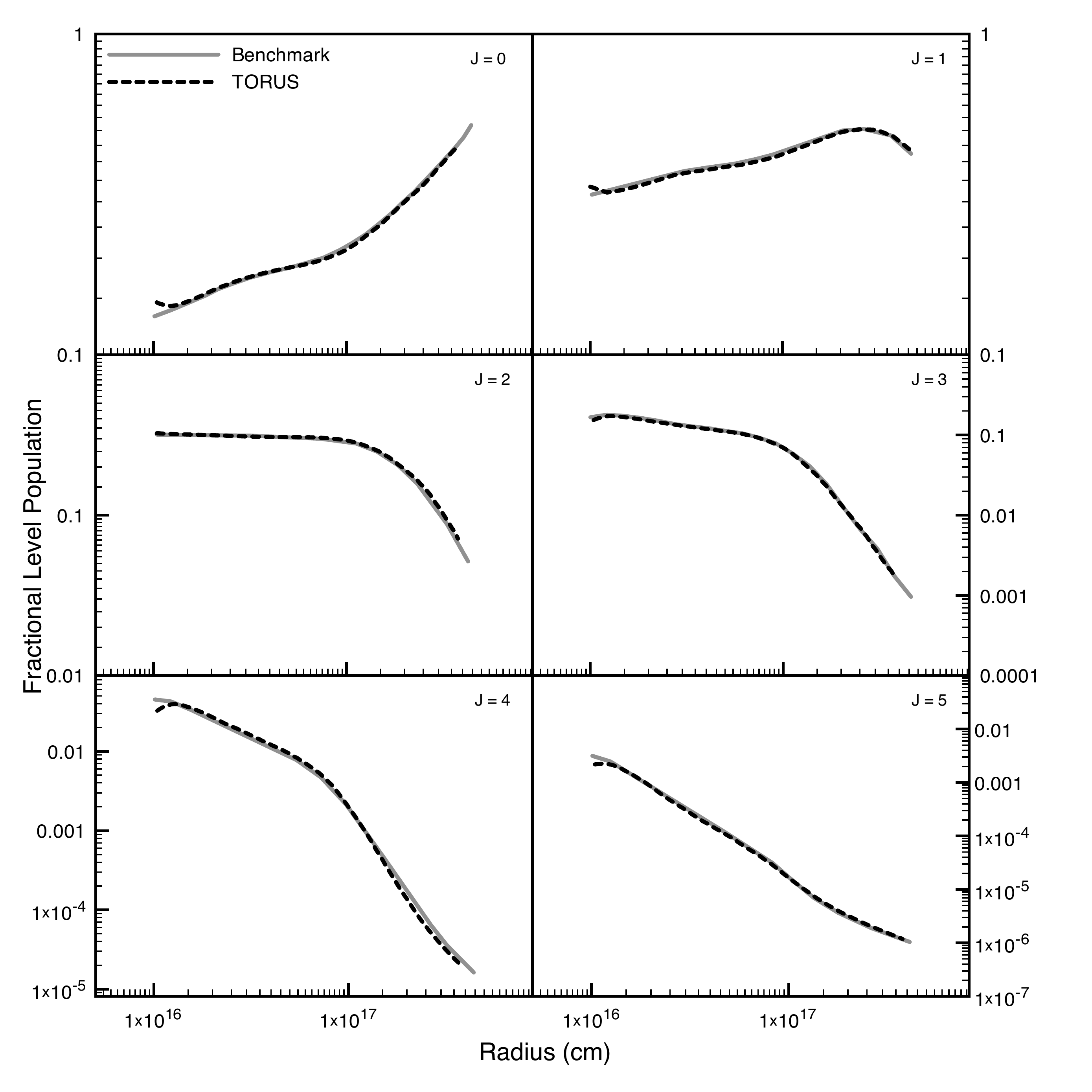}
\caption{
Relative level populations for $J=0$ to $J=5$ for the optically thick benchmark case ($\tau_{J=2-1} \approx 44$) where [HCO$^{+}$] = 10$^{-8}$. Note the deviation from the benchmark of the level populations at the inner boundary reflects differences in the chosen boundary conditions of the models that were combined to produce the benchmark curves.
}
\label{fig:fracpops}
\end{figure*}

The resultant level populations as calculated by \torus are compared for the optically thin scenario (model (a), Figure \ref{fig:fracpops1e-9}) and for the optically thick scenario (model (b), Figure \ref{fig:fracpops}) with an average of other codes (see \citealt{vZ02} for details) that completed this test. The first 8 levels are all converged to better than 0.25 per cent allowing us to be confident that the first 6 are accurate. It is therefore these first 6 levels that are compared. Higher levels are more sparsely populated and are subject to significantly more Monte-Carlo noise. It can be seen that the level populations are consistent with other codes. Without the exact values obtained for each code it is not possible to precisely compare our output with their ensemble average but our code varies by no more than 5 per cent for $J=0$ in the optically thick case. Agreement in the optically thin case is even closer. We note that the average of the results of the codes may not be a good quantitative measure of accuracy but expect that agreement will be closest with other AMC codes.

Using Ng acceleration, the optically thin case converges to better than 1 per cent (RMS all levels) in less than 5 minutes on a single-processor desktop machine (2 GHz Intel Core 2 Duo) having reached 12800 random rays using the ray-doubling method described in section \ref{Implementation}. Within 10 minutes this improves to better than 0.5 per cent using 51200 rays to sample the radiation field. While each level converges at its own rate, in this model the highly populated $J=0$ and $J=1$ levels converge much faster than the other levels having converged to better than 0.05 per cent after 51200 rays.

\begin{figure*}
\center{
  \includegraphics[width=0.8\textwidth]{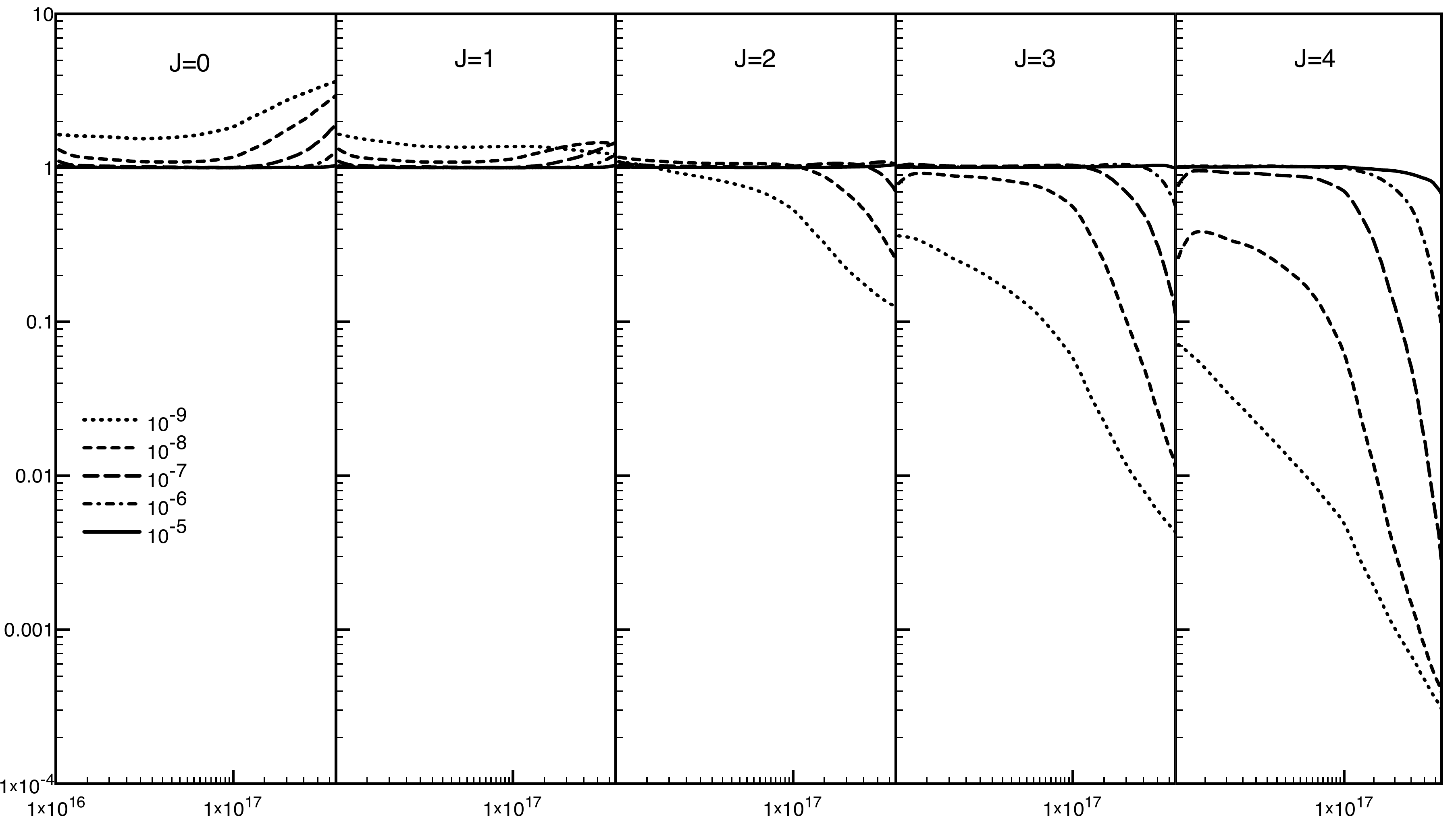}
  \caption{Departure coefficients for 5 different abundances from [HCO$^{+}$] = $10^{-5}$ to $10^{-9}$ (see legend). Both model cases [HCO$^{+}$]  = $10^{-9}$ and [HCO$^{+}$]  = $10^{-8}$ show that they are far from LTE. As expected, for [HCO$^{+}$]  $> 10^{-7}$ it can be seen that the cloud has become sufficiently optically thick in most lines that these levels are well approximated by an LTE solution everywhere except the cloud periphery.}
  \label{fig:dcs}
}
\end{figure*}

The relatively low abundance of HCO$^{+}$ in the model creates a scenario where the cloud is quite optically thin. The (1-0) transition ($\nu_0 = 89.1885$ GHz) has an optical depth of $\leq 5$ from the centre to the edge of the cloud for [HCO$^{+}$] = $10^{-9}$. The critical density for this transition is $\sim 10^5$ cm$^{-3}$. Consequently, throughout most of the cloud radiative processes are dominant over collisional processes therefore the assuming LTE would be inappropriate. Furthermore, the presence of a systematic velocity field allows radiation emitted at the line centre of a particular transition to escape the denser cloud centre more easily, pushing the solution further from LTE. Figure~\ref{fig:dcs} shows the departure coefficients from the LTE case, defined as $\mathbf{n}_i^{nLTE}/\mathbf{n}_i^{LTE}$, the ratio of the calculated relative level population for a specific rotational energy level at some radius to that predicted by assuming the cloud is collision dominated.

\subsection{GG Tau - A raytracing test}
\label{ggtau}

\begin{table}
\centering
\caption{GG Tau model parameters
\label{ggtaumodelparams}}
\begin{tabular}{cll}
\hline
Parameter&Value&(unit)\\
\hline
n$_0$&$6.3 \times 10^{9}$&cm$^{-3}$\\
$^{13}$CO/H$_2$&1.76$\times 10^{-8}$&\\
$r_0$&100&AU\\
$H_0$&14.55&AU\\
$T_0$&30&K\\
$V_0$&3.3&kms$^{-1}$\\
$v_{\textrm{turb}}$&0.2&kms$^{-1}$\\
Inclination&43&$^{\circ}$\\
$r_{\textrm{in}}$&180&AU\\
$r_{\textrm{out}}$&800&AU\\
\hline
\end{tabular}
\end{table}

The \cite*{Dutreyetal94} model of GG Tau, a young multiple star system surrounded by a circumbinary disc, provides an excellent test of the imaging capabilities of \textsc{torus}. The circumbinary disc is assumed to be in LTE and the temperature, H$_2$ number density and velocity profiles are all given in analytical form unlike the tabulated data for the collapsing cloud in section \ref{molebench}. This allows a direct comparison of the predictions of the flux that should be observed from the object in $^{13}$CO by different RT codes. We compare our results with those of Pinte (private communication). A description of the code used to obtain those results is given in \citet{Pinteetal06}. The model parameters and scaling laws common to both simulations are given below and in Table \ref{ggtaumodelparams}. 
\begin{align*}
n(r,z) &= n_0 \left( \frac{r}{r_0} \right) ^{-\frac{11}{4}} \exp\left( \frac{-z}{H(r)} \right)^2 \textrm{cm$^{-3}$}\\
H(r) &= H_0 \left( \frac{r}{r_0} \right) ^{\frac{5}{4}} \textrm{AU}\\
T(r) &= T_0 \left( \frac{r}{r_0} \right) ^{-\frac{1}{2}} \textrm{K} \\
V(r) &= V_0 \left( \frac{r}{r_0} \right) ^{-\frac{1}{2}} \textrm{kms$^{-1}$.}
\end{align*}
As the model is symmetric about the rotational axis, a cylindrical coordinate system $(r,\phi, z)$ is used. \torus takes advantage of rotational symmetry by projecting any coordinate with $\phi \neq 0$ onto the $(r,0,z)$ plane. Naturally, this reduces the number of grid cells required to fill a space to $O(h^2)$ from $O(h^3)$. In order to resolve the disc well it is necessary to devise criteria that ensure the grid is split sufficiently to capture all its features. We used the following conditions to decide whether to split a cell, where $d$ is the cell width and $H(r)$ is the characteristic scaleheight of the disc at radius $r$:
\begin{align*}
\frac{z}{H(r)} < 5  \quad &\textrm{and}  \quad \frac{d}{H(r)} > 0.1\\
\frac{z}{H(r)} \geq 5 \quad & \textrm{and} \quad \frac{z}{d} > 5 \\
r > 0.99 r_{\textrm{in}} \quad & \textrm{and} \quad r < 1.01 r_{\textrm{in}} \quad \textrm{and} \quad d > 0.1r_{\textrm{in}}
\end{align*}

If any of these conditions were true then the cell should be split. The cell is not split if $r < 0.99 r_{\textrm{in}}$ or $r > 1.01 r_{\textrm{in}}$ even if another condition is satisfied as the model is not defined outside these radii. The minimum cell depth is 3 and the maximum depth is set to 20, although this is never reached. The first two conditions cover the entire vertical extent of the disc and stipulate that at least 50 cells must be used to cover the first 5 scaleheights above the disc midplane. Beyond this, where the disc is far more tenuous, the criterion is relaxed so that the maximum cell size is less than 20 per cent of the $z-$coordinate of the cell. The third criterion ensures the adequate resolution of the inner-edge of the disc.

\begin{figure}
  \includegraphics[width=0.475 \textwidth]{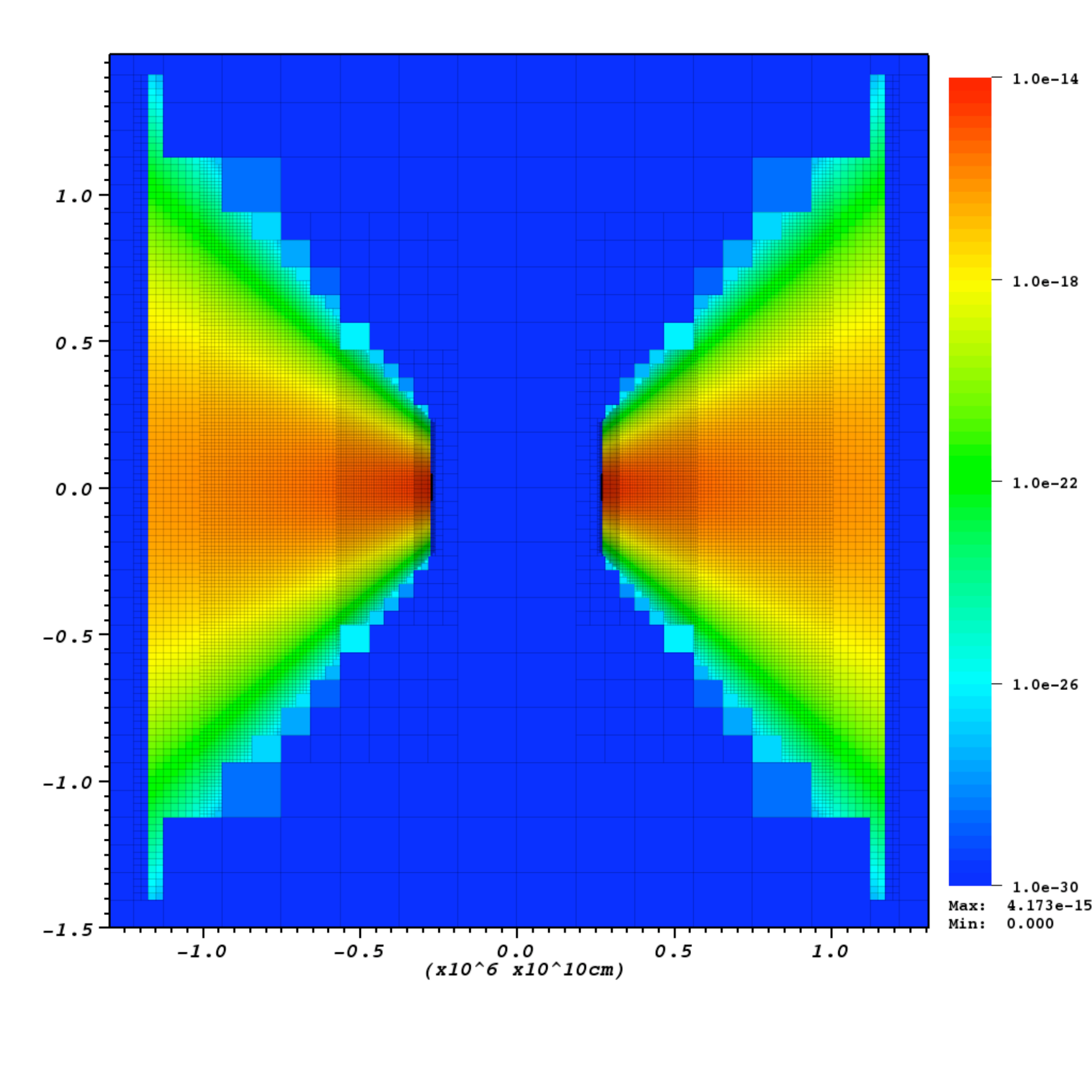}
  \caption{The discretised density profile of GG Tau in g cm$^{-3}$. The log scale captures the steep decline in material away from the disc midplane.}
  \label{fig:ggtaurho}
\end{figure}

\begin{figure}
  \includegraphics[width=0.475 \textwidth]{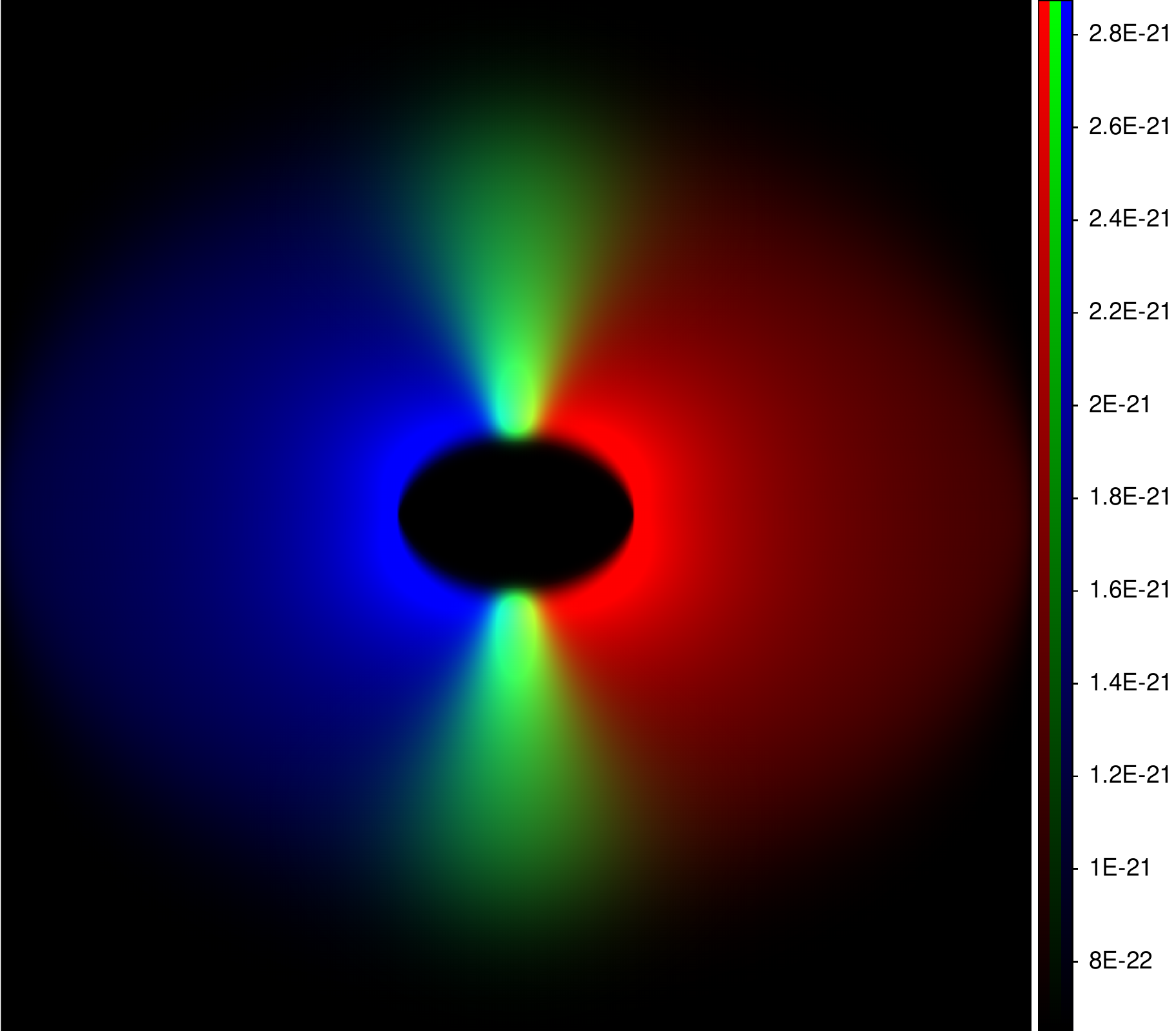}
  \caption{This three colour plot of GG Tau in \low~(1-0) shows that the observer is inclined to the midplane (as evidenced by the elliptical evacuated region in the centre of the disc) and that the disc is relatively optically thin almost throughout. The red component is material moving away from the obsever, blue towards and green has no radial velocity component. The intensity is given in erg s$^{-1}$ cm$^{-2}$ sr$^{-1}$ Hz$^{-1}$.}
  \label{fig:ggtau3col}
\end{figure}

When the disc is split according to these criteria, a cylindrical region with diameter of $6 \times 10^{16} \textrm{cm} \approx 2000\ \textrm{AU}$ is split into 23,782 cells, the smallest of which is 0.12 AU across. Figure~\ref{fig:ggtaurho} shows the discretized density profile and Figure~\ref{fig:ggtau3col} shows the disc imaged in \low~(1-0).

\begin{figure}
  \includegraphics[width=0.475\textwidth]{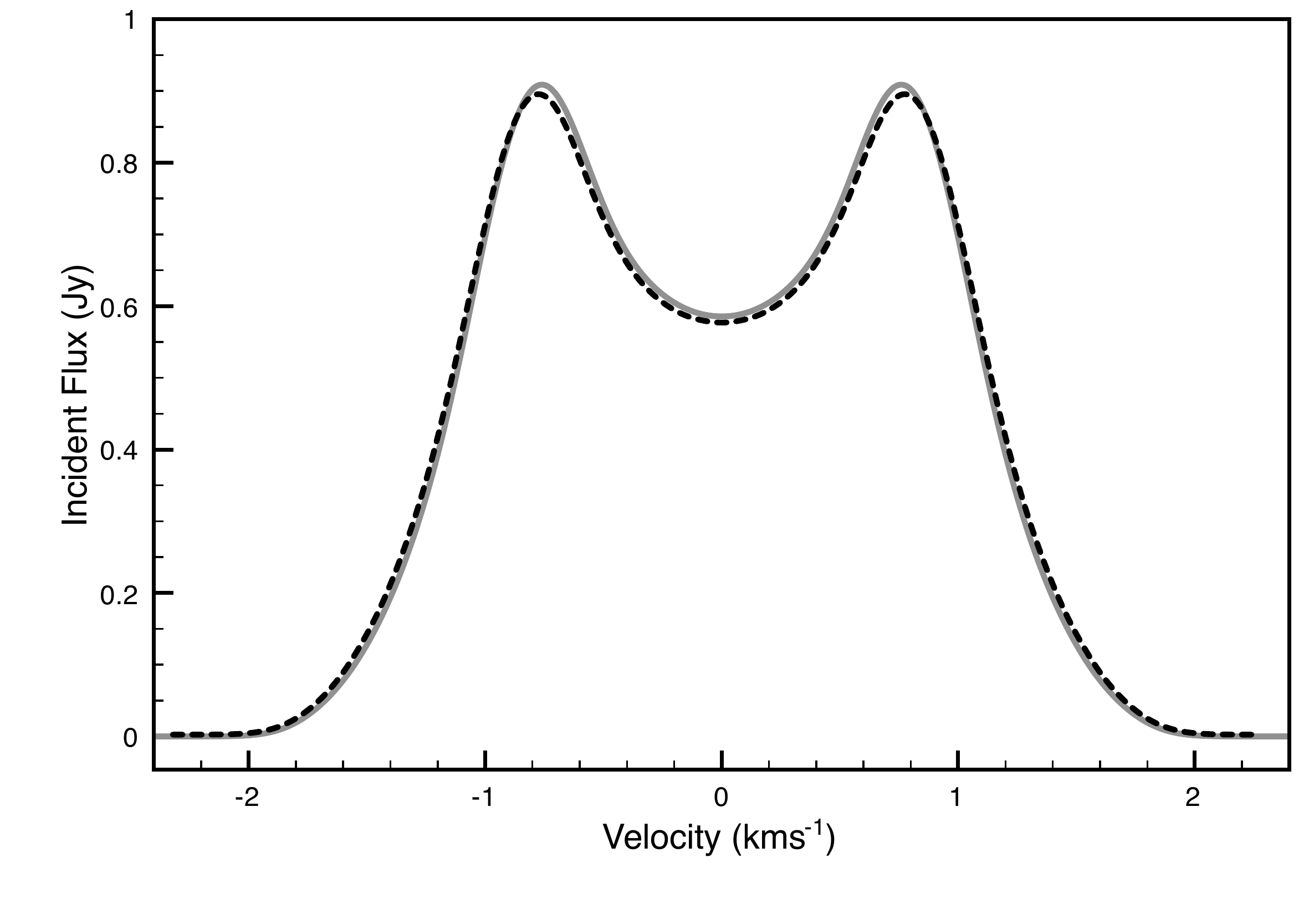}
  \caption{This line profile shows the excellent agreement between \textsc{mcfost} (solid grey line) and \torus (dashed black line). A distance of 150 parsecs was assumed to convert intensity into flux. The agreement is well within observational error and discrepancies may be put down to minor differences in discretization strategy. This confirms that the \torus raytracing routines are sound.}
  \label{fig:ggtaucomparison}
\end{figure}

Figure~\ref{fig:ggtaucomparison} illustrates the excellent agreement between \textsc{mcfost} and \torus line profiles in this  test case. The geometry and simplifying assumptions make it possible to test the capabilities of the rendering code. Having passed both tests we are confident in attempting a far more complex geometry that does not exhibit any symmetry and where no parameter is defined analytically.
\section{An efficient particle-mesh algorithm}
\label{sphtogrid}

In order to map the density structure of a particle-based representation onto the adaptive mesh employed by \torus it is necessary to use a particle-mesh algorithm that interpolates data stored at irregularly spaced points onto the grid. Many  methods exist that already do this; the simplest is to use only the particles within a cell to determine a parameter value. For example, the density is taken to be the average density over all the particles within the cell or the sum of the particle masses divided by the cell volume \citep{Kurosawaetal04}. However, the variation between these two calculations may be large and neither takes into account the concept of smoothing inherent in the SPH technique. Furthermore, whilst this technique tends to give an accurate result in high density regions, where the particle density is naturally greatest, it is not capable of giving an answer in regions where no particles exist and further provision must be made to evaluate parameters in empty cells. More sophisticated algorithms take into account a particle's contribution to its associated cell and its nearest neighbours using a linear or quadratic kernel, namely the cloud-in-cell algorithm, however even this algorithm fails to take into account the variable radius over which each individual particle contributes to its environs. In fact, in order to accurately determine information about a parameter at any given point in space from a particle ensemble, it is necessary, in theory, to calculate and sum the contribution from every particle.
In practice, to make SPH calculations tractable, each particle is assigned a characteristic size scale that defines its `sphere of influence', beyond which the contribution is presumed to be negligible. This variable smoothing length, $h_i$, for a particle with index $i$, is either determined by 
\begin{equation}
\label{eqn:h}
h_i = \eta {\left(\frac{m_i}{\rho_i}\right)}^{1/\nu},
\end{equation}
where $\eta$ is a constant controlling the approximate number of nearest neighbours a particle has and $\nu$ is the dimensionality of the simulation \citep{SpringelandHern02, PriceandMon04, PriceandMon07}, or is determined at each timestep to control exactly the number of nearest neighbours \citep{BaBoBromm03}, depending on the desired resolution of the calculation. Note that in the latter method, no closed form exists for $h_i$. The weight of each particle's contribution is determined by the smoothing kernel; a normalized analytical function of the particle-point separation in units of smoothing lengths, $W(|\mathbf{P}-\mathbf{p_i}|,h_i)$. Throughout this paper we use the cubic spline kernel (see equation \ref{eq:kernel}). This kernel is popular due to its ease of computation and compact support over $2h$. Kernels are discussed in greater detail in \cite{Mon92}. Below we outline the method we use to map the information stored in the particle representation onto the AMR grid.

\subsection{Creating the grid}

An AMR grid is created by the repeated subdivision of an initial unrefined cell centred on the entire region of interest. The parent is bisected once in each dimension, thus each child has a volume $2^{-\nu}$ times that of its parent. As we are dealing with complex astrophysical structures in this section we use a full 3-dimensional representation of the space so each parent has 8 child cells, or octals. More detail of the AMR method is given in \cite{Harries00} and \cite{Symington05a}.

Problems in star formation typically span many orders of magnitude in both space and density. SPH and AMR naturally resolve the fine structure of these problems well whereas a fixed, regular grid will often be insufficient to capture crucial information like peaks in the density profile where a protostellar core far smaller than the size of a grid cell has begun to accrete material. 
The flexibility of AMR is that it will adapt the resolution given to a region depending on the criteria the user applies. Typically, one is interested in resolving the temperature and density profile in a region, although when calculating line radiative transfer accurately it is necessary to accurately resolve variations in the velocity field as well owing to the anisotropy in the absorption introduced by the line profile function. To create the grid we test each parent cell against a mass criterion, a density criterion and a velocity criterion. Where these criteria are not met, a cell is recursively split until the conditions in each child cell are met up to a specified maximum depth of recursion.

The mass per cell criterion determines the maximum number of particles that may occupy a cell, thus ensuring that no cell has more than a certain fraction of the entire mass of the simulation within it. Typically, this value is set to $\sim 50$ particles; this balances the need for accurate determination of densities within cells and sufficient grid resolution within memory constraints.

The density criterion dictates the range of particle densities within a cell. This criterion facilitates the resolution of density gradients which are both physically and computationally important in radiative transfer calculations (e.g. treatment of shocks, jets, etc.). Large changes in density from cell to cell can have deleterious effects on the convergence of the iterative algorithm used to determine the relative level populations of a molecular species within the cell so it is crucial that the volume over which they change is well resolved. Typically, this criterion is set so that the maximum density stored on a particle within the cell is not greater than twice that of the minimum density.

The velocity criterion regulates the variation in the magnitude of the velocity within a cell. Again, it is vital to have good velocity resolution over a cell to obtain the correct line profile shape since absorption and emission of line radiation is a strong function of Doppler-shifted frequency. Therefore, where possible we use this criterion to ensure that the range of velocities within a cell is less than 5 times that of the turbulent line velocity (see equation \ref{eqn:microturb}).

\begin{figure}
\includegraphics[width = 0.475 \textwidth]{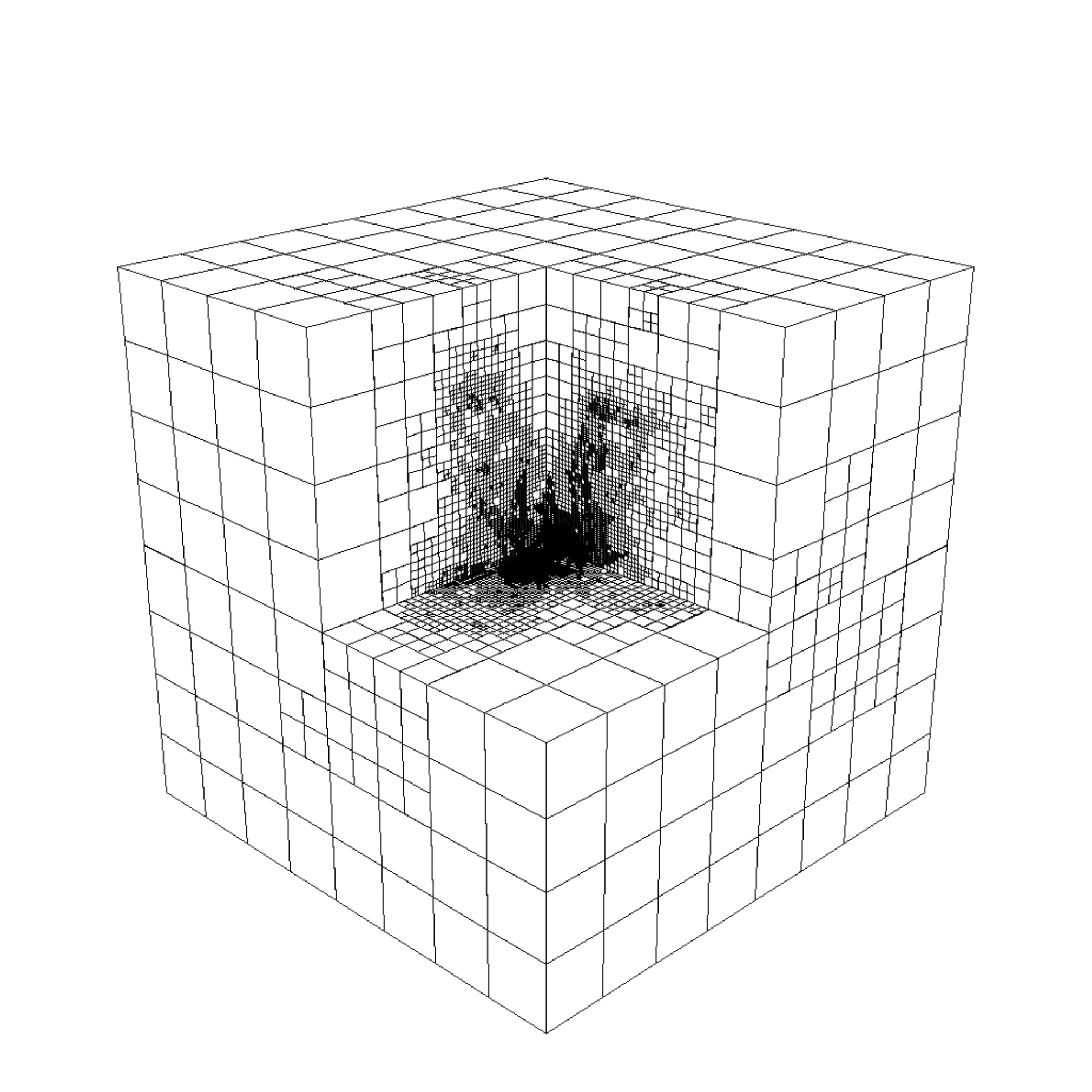}
  \caption{A cut-through of an adaptively refined grid. The minimum AMR depth of 3 can clearly be seen here giving a coarse grid of $8^{3}$ cells; the deepest level of recursion is 23. The spatial extent of this grid is $5 \times 10^{18}$ cm.}
  \label{fig:meshplot}
\end{figure}

If any of the conditions in the parent cell meet the splitting criteria then the cell is split. This process is iterated until each cell no longer triggers any of the criteria or a maximum cell depth is reached. This is to ensure that the simulation does not run out of memory. We impose a further condition that the minimum subcell size is no smaller than the smallest inter-particle separation ($\sim 0.25 h_{\rm{min}}$), which is commensurate with the natural resolution of the SPH simulation. Figure~\ref{fig:meshplot} shows a cut-through version of a mesh created using the above splitting criteria. The minimum cell size in this grid is $\sim 0.05$ AU. 

\subsection{Determining cell parameters}

Having created a grid it is then necessary to populate each cell with parameters pertinent to the calculation that is to be performed. The method outlined here is optimized to take advantage of geometry where possible but is not prejudiced against the general case where the distribution of particles is not known \textit{a priori}.

For each cell in the grid, the centre is chosen to be the point in space where the conditions that pervade the  cell are to be determined. Whilst it is trivial to determine the region over which a particle acts (because the distance at which a particle ceases to contribute is fixed at 2$h_i$), the inverse problem of determining which particles contribute to the grid cell centre is less trivial. That is, some may lie within the cell but some may not. Furthermore, some may lie within the cell yet not contribute to it at all. Na\"{i}vely, one might assume that by finding the particle with the greatest smoothing length, \hmax and subsequently checking for all particles that lie within 2\hmax of a point one is guaranteed to recover the most accurate answer possible. Whilst this is certainly true, it turns out to be so inefficient that determining the contents of an entire grid like this would take an impractically long time. Not only does \hmax necessarily overestimate the smoothing lengths of all but one particle in the ensemble, but also, due to the direct relationship between smoothing length and density it is unrepresentative of the entire physical region; very low density volumes being the exception rather than the rule in circumstellar discs and star-forming clusters. Moreover, these very low density regions play a negligible role in radiative transfer and are naturally undersampled by the SPH technique. We define the parameter $h_{95\%}$ as the smoothing length of the particle lying at the 95$^{\textrm{th}}$ percentile of the distribution of smoothing lengths of all particles. This length is often as much as an order of magnitude less than that of \hmax but still greater than that of all but the 5 per cent of particles representing the least well-populated areas of the physical space. However, as the adaptive mesh is already split so that the cell size is commensurate with the smoothing lengths of the particles it contains, a better estimate for the critical lengthscale over which to search for nearby particles is \rcrit = $\min(4d, h_{95\%})$, where $d$ is the cell width. The factor of 4 has been empirically derived as the smallest factor that retains the furthestmost particles that are likely to contribute whilst keeping computational speed high. Equation \ref{eqn:h} shows that for each ten-fold reduction in $r$, 3 orders of magnitude of density resolution are lost. However, as the radius over which contributing particles are searched for reduces, the volume (and hence the computational effort) reduces as $r^3$, making the problem more accessible. Furthermore, as it is the least dense regions that are no longer resolved, the impact on the solution is minimised. If however there are no particles in a point's vicinity then the code makes another attempt to locate a nearby particle that can be used to derive the local density. The algorithm is repeated with a wider search volume whose radius is determined by
\begin{equation}
\mrcrit = \min(\max(4d,2h_{95\%}),0.5h_{\textrm{max}}).
\end{equation}
This second condition covers almost all cases where the initial condition does not suffice but if even this radius is not sufficient then one last attempt is made using $r_{\textrm{crit}} = h_{\textrm{max}}$; beyond this the cell is justifiably declared empty and its parameters set to some global minimum.

Sorting the particle list by ascending $x-$coordinate facilitates the reduction of the number of candidates that can contribute to a point, \textbf{P}. This is achieved by removing all particles that lie outside 2\rcrit in the abscissal dimension, leaving only particles whose $x-$coordinate lies within $\textbf{P}_x\pm \mrcrit$. In the case where all the particles are equally spaced this reduces the number of comparisons necessary to $4\sqrt[3]{N_{\textrm{part}}}$. Each particle in this reduced list is then tested against the same criterion in $y$, i.e. $|{\textbf{P}^y - \textbf{p}_i^y}|< 2r_{\textrm{crit}}$. For all those particles where this criterion is satisfied a further test for the $z-$component is made. This process discovers all particles that lie within a cube of side length 4\rcrit centred about the point. Whilst this means that only approximately half the particles are expected to lie within a sphere of radius 2\rcrit, it is still more efficient than directly calculating the pairwise distance for each particle in the entire ensemble. The logic is such that, at each step in the algorithm, it is necessary to do less work than the stage before. Furthermore, the order in which the criteria are applied has been selected in such a way as to reject the greatest number of particles as early on in the process as possible; that is, culling by $x$ then $y$ then $z$ is the most efficient method of locating nearby particles for a disc with its semi-major axis in the $xy$-plane.
Once the reduced list of particles has been created it is then a matter of calculating the contribution of each particle to the point.

For each remaining particle, the contribution to the point is determined by the kernel smoothing function, $W$. A common smoothing kernel is the cubic spline, defined below,
\begin{equation}
W(q,h) = \frac{\mathcal{W}(q)}{h^{\nu}} = \frac{\sigma}{h^{\nu}}
\left\{
\begin{array}{l l}
    1-\frac{3}{2}q^{2}+\frac{3}{4}q^{3} & \rm{for}  \ 0 \leq q <  1, \\[1.25ex]
    \frac{1}{4}(2-q)^{3} & \rm{for} \ 1 \leq q < 2, \\[1.25ex]
    0 & \rm{otherwise},
\end{array}
\right.
\label{eq:kernel}
\end{equation}
where
\begin{equation}
q_i = \frac{|{\textbf{P} - \textbf{p}_i}|}{h_i}
\end{equation}
and $\sigma$ is the normalization constant for the dimensionless kernel $\mathcal{W}(q)$ defined such that $\int_0^{\infty} \mathcal{W}(q)dV = 1$. For $\nu~=~3$, $\sigma = 1/\pi$. By utilizing the property of compact support possessed by the kernel smoothing function it is possible to discard those particles with $q \geq 2$ without having to evaluate the kernel function.

In a homogeneous, isotropic distribution of 10$^6$ particles, this smoothing kernel reduces the number of particles able to contribute to a point by approximately five orders of magnitude. Specifically, the number of particles able to contribute is independent of volume or number, being constant at $\sim 60$ for $\eta = 1.2$.

In the SPH method, the value of a scalar parameter at a point in space, A($\mathbf{r}$), is the weighted sum of the stored parameter values, $A_i$, over all nearby particles, i.e.
\begin{align}
\label{eqn:kernelweight}
A(\mathbf{r}) &= \sum^{N_{\textrm{part}}}_{1}{\frac{m_i}{\rho_ih^{\nu}} \mathcal{W}(q_i) A_{i}}\\
&= \eta^{-\nu}\sum^{N_{\textrm{part}}}_{1}\mathcal{W}(q_i)A_{i}, \qquad \qquad \quad \textrm{by equation \ref{eqn:h}} 
\end{align}
This formulation assumes that $\sum{\mathcal{W}(q_i)} \approx 1$ which is typically true for $\mathbf{r}$ in the bulk of the simulation surrounded by a full complement of neighbours. Where it is true it is possible to normalize $A(\mathbf{r})$ by dividing by $\sum{\mathcal{W}(q_i)} \approx 1$, reducing the variation associated with kernel smoothing. The superposition of many spherically bound kernels on to a regular 3D grid will not recover a constant field but rather it will tend to oscillate around the true value; the field will be over-estimated at the particle position (where the weighting function at that point is 1 and the sum over all other particles is positive) and will similarly be underestimated at the point in space centred at the midpoint of surrounding kernels. This normalization can certainly improve the appearance of an image, (see \citealt{Price07}) but can cause problems at free surfaces where a smooth decay is preferred.

Particles with $h > \sqrt[3]{V_{\textrm{sim}}/N_{\textrm{part}}}$, the average linear separation of particles in the simulation volume, $V_\textrm{sim}$, are expected to lie close to a free surface and are initially classed as `hull' particles. The algorithm adaptively decides whether to normalize or not depending upon one of two strategies: either, if $\sum{\mathcal{W}(q_i)} > 0.3$ then normalize; or normalize only if all contributing particles are determined to be `bulk' particles, (as opposed to hull particles). Either strategy relies upon empirically determined parameters, $\sum{\mathcal{W}(q_i)} > 0.3$ typically recovers total masses most similar to the SPH simulation.  The designation of a hull particle may change if it is subsequently found to belong to a set of particles where $\sum{\mathcal{W}(q_i)} > 0.3$, whereby it is classed again as a bulk particle. Similarly, it is possible, though unlikely, that a bulk particle may become a hull particle if it belongs to a set that fails to meet the criterion. In any case, it is possible for a user to override this strategy as it is not obviously desirable to let the temperature or velocity decay at a free surface even in the apparent absence of matter.

Once the contributing particles have been located and their weights determined they can be reused for each parameter that needs to be found. A vector such as velocity must be split into its 3 components and each one must be determined before outputting the resultant vector. The algorithm is repeated for each point until the grid has been filled with data.
 
\section{Imaging simulated star formation regions}
\label{sph}
We have taken snapshots of an SPH star formation simulation by \cite{BaBoBromm03}. The simulation used $3.5 \times 10^{6}$ particles to trace the evolution of a 50 $M_{\odot}$ molecular cloud. The cloud was initially spherically symmetric with a uniform density and temperature (10 K). A supersonic turbulent velocity field ($\mathcal{M} = 6.4$) was imposed on the cloud to create anisotropies which ultimately led to the formation of a dense pre-stellar core after approximately 1 free-fall time. Here, $t_{\textrm{ff}} \approx 1.9 \times 10^{5}$ yr. \citeauthor{BaBoBromm03} observed that high density regions form from converging gas flows during the decay of the turbulent velocity field. These enhancements continue to increase in density until a particle exceeds a density of $10^{-11}$g cm$^{-3}$, when a sink particle is created with a mass   to the sum of the masses of all particles within a 5~AU radius. The sink particle inherits a weighted average of all the properties of the contributing particles and they are destroyed. This ameliorates the computational slowdown caused by resolving the acceleration caused by large forces generated by particles with small separations which requires a very small timestep.

During the early stages of collapse, the relatively tenuous gas allows energy to be radiated with ease. At this stage the gas was assumed to be isothermal. If the rate at which energy is released exceeds the rate at which the gas can cool then the local environment begins to heat rapidly. \cite{Masunaga00} showed that the gas begins to heat significantly at a density of $\rho > 10^{-13}$ g cm$^{-3}$ depending on gas opacity (and initial temperature). To avoid performing a computationally expensive full radiative treatment, \citeauthor{BaBoBromm03} used a barotropic equation of state $p = K\rho^{\eta}$. The barotropic exponent, $\eta$, equals unity where the gas is isothermal (for $\rho \leq 10^{-13}$ g cm$^{-3}$) increases to 7/5 for $\rho > 10^{-13}$ g cm$^{-3}$. This equates to a temperature-density relationship of
\begin{equation}
\label{eqn:Tgas}
T_{\textrm{gas}} = \max\left(10,10\left(\frac{\rho}{10^{-13}}\right)^{2/5}\right).
\end{equation}

The frames we have used in this paper are taken at 1.0, 1.1, 1.2, 1.3 and 1.4 $t_{\textrm{ff}}$; the first snapshot occurs just before the start of star formation, the second is taken during the first star formation episode, while the third and forth are taken during an interstitial period during which a second and third star-forming core is formed. The final snapshot is taken at the end of the simulation when over 50 objects have been formed, some of which are still accreting. At the end of the simulation, approximately 1/5 of the SPH gas particles ($\approx$ 10 M$_{\odot}$) have been accreted onto the sink particles.

\subsection{Transposing SPH particles onto the AMR grid}
\label{discretisation}

\begin{figure}
\includegraphics[width = 0.475 \textwidth]{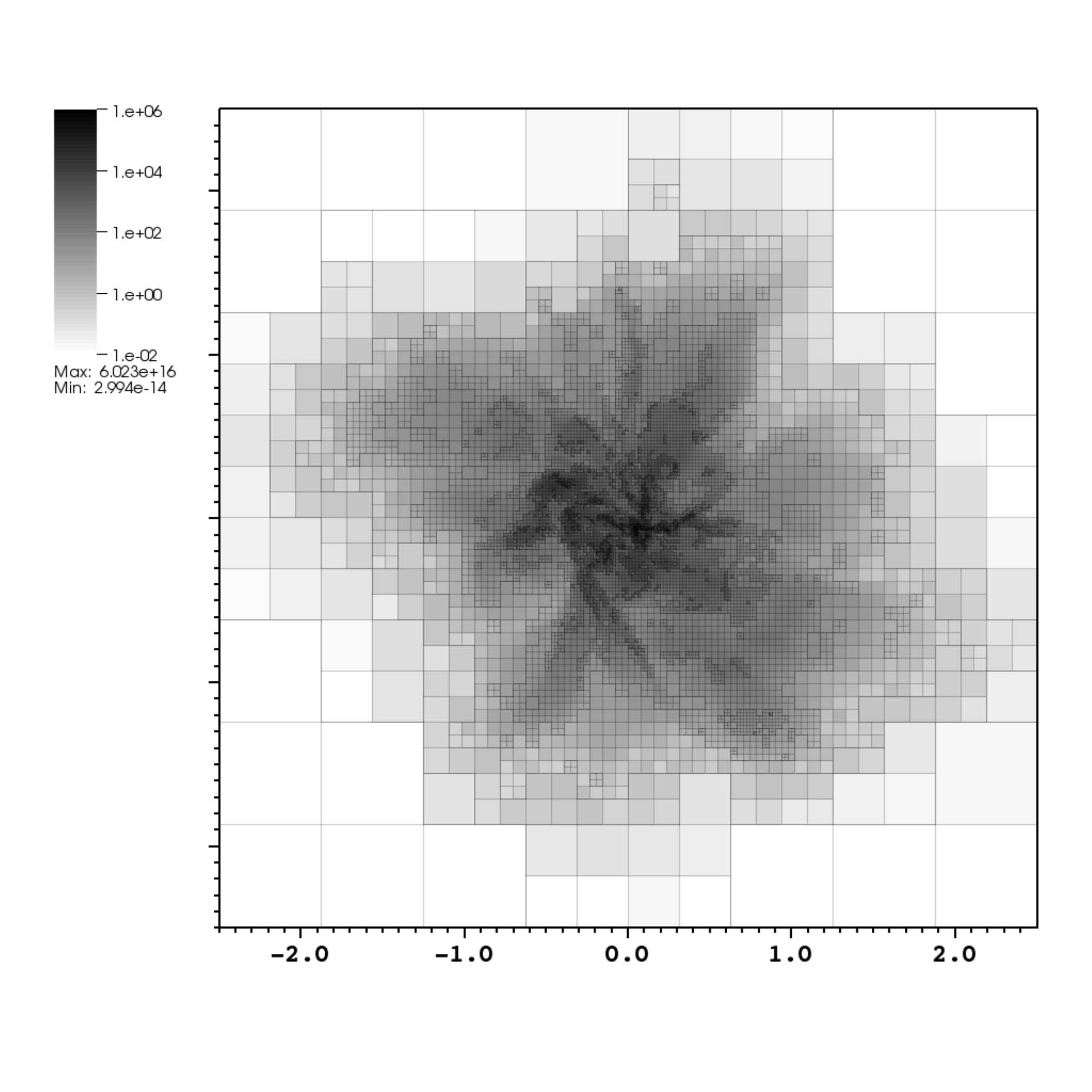}
  \caption{A slice through the $xy-$plane illustrating the dynamic range of densities that can be captured within the cloud using AMR. Thirty orders of magnitude are present in the simulation with densities reaching as high as $2\times10^{-7}$g cm$^{-3}$ and as low as $5 \times 10^{-36}$g cm$^{-3}$ (although very little material by mass is actually present below $\sim 10^{-25}$g cm $^{-3}$). The greyscale for this plot is given in $n$(H$_2$).}
  \label{fig:rhoplot}
\end{figure}

We discretized each SPH snapshot using the algorithm outlined in section \ref{sphtogrid} using the same splitting criteria for each timestep; no more than 50 particles were allowed within one cell, the ratio of the maximum density to the minimum density must be no greater than 2 and the range of velocity magnitudes must be less than 5$v_{\textrm{turb}}$. Figure~\ref{fig:rhoplot} illustrates how an SPH representation of density can be discretized onto a mesh.

The process of discretization will necessarily introduce some further uncertainty in the exact distribution of mass throughout the cloud, however the algorithm has been designed to minimize any systematic error which may arise through using too coarse a grid to represent the space, a deeper study of which is undertaken in \cite{AcrHarRun10}. The algorithm has been successfully used on galaxies \citep{AcrDouDob10}, circumstellar discs \citep{AcrHarRun10}, star-forming clusters (this work) and as a crude test of discretization accuracy the total mass stored on the grid is found to lie within 5 per cent of the true mass in all cases.
\begin{figure}
\includegraphics[width = 0.475 \textwidth]{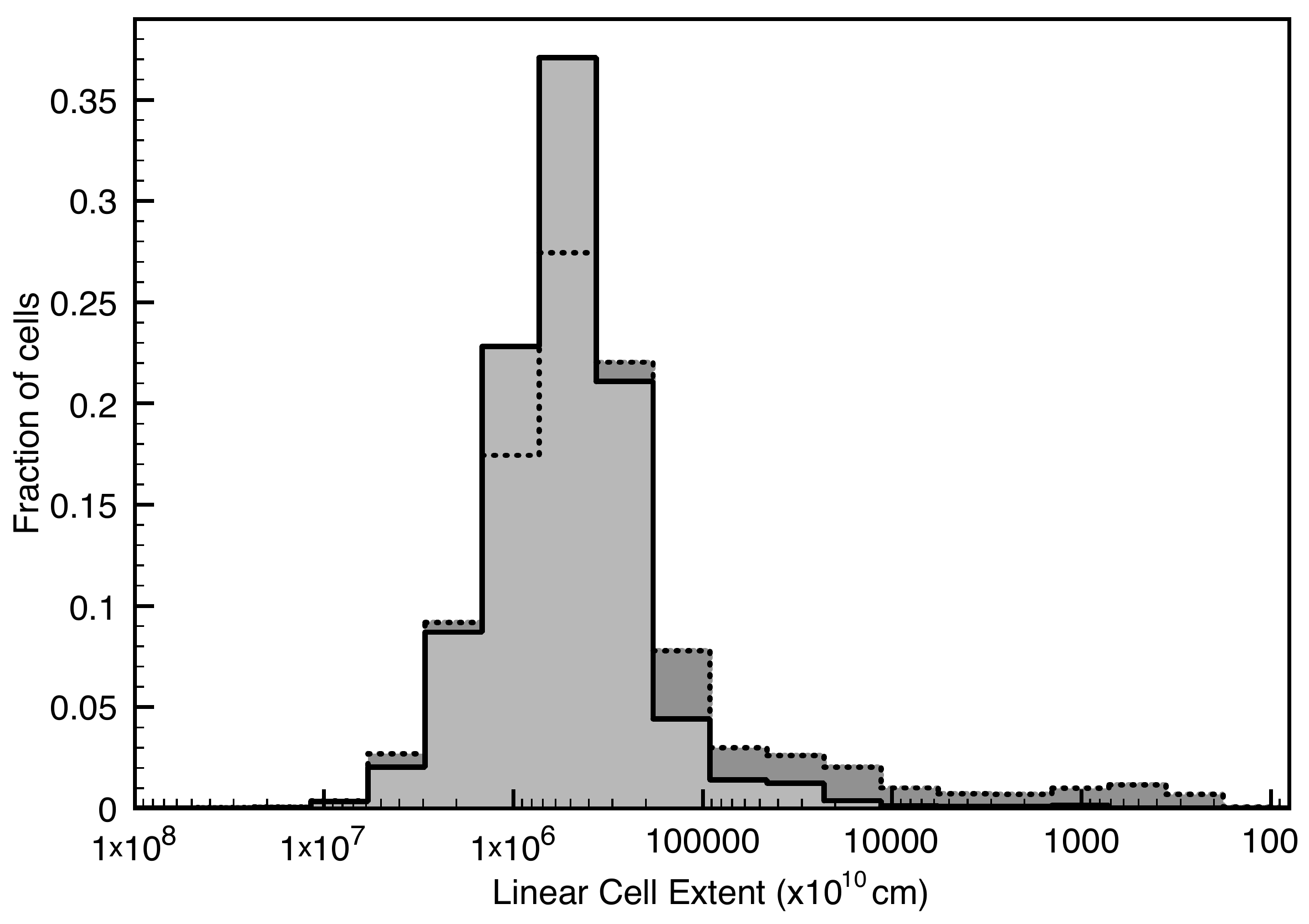}
 \caption{Histogram of the relative fraction of AMR cells (dark grey) at each depth for $t=1.1t_{\textrm{ff}}$. The smoothing lengths of the SPH particles (light grey) were binned according to their closest association in log-space. Note that the x-axis is reversed to show the proportion of cells at lower depths on the left.}
 \label{fig:AMRvsSPH}
\end{figure}
In this work, the mass discretization error is better than 3 per cent and in the absence of overdense regions, better than 1 per cent (see Table \ref{table:sphtoamr}). Furthermore, a comparison of the number of cells in each level of refinement against a histogram of the smoothing lengths of the particles (Figure~\ref{fig:AMRvsSPH}) shows broad agreement, except in the highly refined tail of the AMR cell distribution. This may be because of the additional constraints required for adequate radiative transfer, additional velocity splitting in particular. Over all 5 frames, the maximum level of refinement was 22, resulting in the smallest octal having a linear dimension of $\sim$ 0.08~AU, commensurate with the shortest inter-particle spacing. On average, this discretization policy resulted in a ratio of particles:cells of approximately 1/3. 
\begin{table*}
	\caption{Table of SPH properties and corresponding AMR grid properties for each timestep.}
	\label{table:sphtoamr}
\begin{tabular}{ccccccccc}

  &\multicolumn{ 3}{c}{SPH properties} & \multicolumn{ 3}{c}{Grid properties} &            & \\
           \hline
   Timestep & N$_{\textrm{part}}$ & N$_{\textrm{pt}}$ & $\rho_{\textrm{max}}$ & N$_{\textrm{octals}}$ & Max Depth & Mass (M$_{\odot}$) & Mass Error (per cent)& N$_{\textrm{part}}$/N$_{\textrm{octals}}$ \\
 \hline          
  1.0 \tff & 3500000 & 0  & 2.18$\times 10^{-16}$ & 1097622 & 14 & 50.43 & -0.850 & 3.19 \\
  1.1 \tff & 3455362 & 8  & 7.65$\times 10^{-10}$ & 1118601 & 21 & 50.18 & -0.369 & 3.09 \\
  1.2 \tff & 3287092 & 18 & 2.49$\times 10^{-09}$ & 1112357 & 23 & 49.42 & 1.16 & 2.96 \\
  1.3 \tff & 3248783 & 27 & 2.92$\times 10^{-09}$ & 1174818 & 21 & 49.14 & 1.73 & 2.77 \\
  1.4 \tff & 3087719 & 50 & 3.17$\times 10^{-09}$ & 1151781 & 20 & 48.67 & 2.66 & 2.68 \\
  \hline
\end{tabular}
\end{table*}

Our analysis of this molecular cloud seeks to examine the conclusions of \cite{Ayliffeetal07} following the analysis of \cite{Walshetal04}. Accordingly we use the \low~(1-0) transition ($\nu_0 = 110.2013542798$ GHz) to trace low density envelopes. Initially, we assume an abundance of [$^{13}$CO] = $10^{-6}$ relative to H$_{2}$. To trace gas closer in to the cores within the cluster we use the C$^{18}$O (1-0) transition ($\nu_0 = 109.7821734$ GHz, [C$^{18}$O] = $10^{-7}$) and to define cores we use the \nnhs(1-0) transition ($\nu_0= 93.1737$ GHz) because of its high critical density ($n_{\mathrm{crit}} \approx 1.4 \times 10^5$). For \nnhs we assume a constant relative abundance of $1.5 \times 10^{-10}$, in line with values obtained for well-studied cores in the Taurus-Auriga molecular complex \citep{Tafallaetal04}. For all molecules, we assume a non-thermal turbulence parameter with a fullwidth of 0.3 km s$^{-1} \sim c_s$ for H$_2$ at the simulated temperatures.

We superimposed the \low, \highs and \nnhs molecular data on to the discretized grid for a total of 15 distinct grids (3 molecules and 5 timesteps). We derived a non-LTE solution for each grid using the molecular line transfer code outlined in section \ref{Implementation}. Table \ref{table:gridconvergence} shows the mean error of the least well converged level populations ($J=4$ in all cases) over all grid cells. A further condition that every cell had to have an RMS error of less than 1 per cent in $J=0$ and $J=1$ was imposed to reduce any pixel-to-pixel variance in intensity in the subsequent raytracing of the grid.

\begin{table}
	\caption{Fractional error (in per cent) of the least well converged level in each grid for each timestep.}
	\label{table:gridconvergence}
	\centering
\begin{tabular}{llll}
          &\multicolumn{ 3}{c}{Species} \\
          \hline
          & \low & \high & \nnh \\
           \hline
  1.0 \tff & 0.30& 0.14& 1.03  \\
  1.1 \tff & 0.27 & 0.19& 1.98  \\
  1.2 \tff & 0.25 & 0.07& 0.54  \\
  1.3 \tff & 0.16&  0.12& 0.59  \\
  1.4 \tff & 0.14&  0.15&  0.71 \\
\hline
\end{tabular}  
\end{table}

\begin{figure}
\includegraphics[width = 0.475 \textwidth]{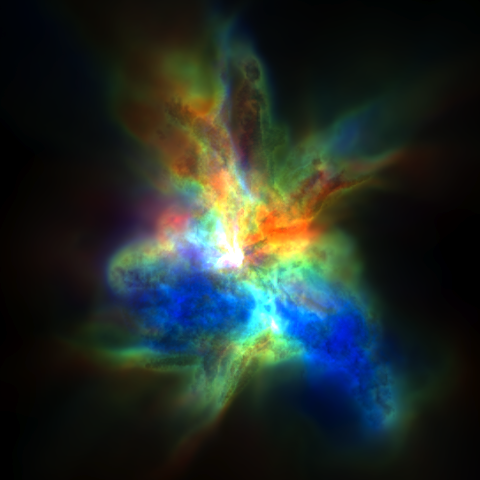}
  \caption{A cluster observed in HCO$^{+}$ (1-0). Blue represents material moving towards from the observer; green, stationary material and red receding gas. The smooth transitions in intensity are due to the careful choice of normalization strategy, adaptive pixel sampling and density subsampling.}
  \label{fig:clusterrgb}
\end{figure}

To improve the statistics, each of the 15 grids were observed from 20 equally spaced positions on a sphere with radius 140 pc. Care was taken to maximize the separation of each observer position giving equal coverage of the viewing sphere from each perspective. A $1024 \times 1024 \times 80$ element datacube was created for each observer position-tracer-timestep triplet. This work therefore contains 300 distinct datasets upon which to perform statistical analysis. The maps have a linear spatial resolution of $2 \times 10^{15}$ cm or $\sim$ 133AU per element. This is equivalent to a spatial resolution of $\sim$ 1 arcsec at 140pc. From the known velocity distribution of the SPH particles, we expect emission to come from a range of velocities within 3.2 km s$^{-1}$  of the line centre giving a velocity resolution of 0.08 km s$^{-1}$. Figure~\ref{fig:clusterrgb} shows an example image that can be produced by combining the non-LTE line transfer algorithm, the SPH particle interpolation and the raytracing routine.
In these grids, molecular abundance has been assumed to remain constant in both time and space. Without a detailed evolution history, it is not possible to predict with any certainty the local chemistry in any region of the grid. In section \ref{chemistry} we describe a simple chemical model to take into account the expected CO freezeout in the cloud. In the absence of better chemical modelling, we assume that the abundance of \nnhs remains constant, which is appropriate given the age of the cluster \citep{Bergin97}.

\subsection{Chemistry}
\label{chemistry}
Molecular adsorption onto dust grain surfaces is known to be an effective depletion mechanism in many carbonaceous species at densities of $n$(H$_2$) = $3\times 10^4$ cm$^{-3}$ \citep{Bacmannetal02}. Notable absentees in this list are N-containing species including \nnh. \highs is expected to trace a larger scale than \nnhs not only because of its lower critical density but also because it effectively freezes out at densities above $\sim 3\times 10^4$ cm$^{-3}$. Therefore, in the cluster, \highs intensity is weighted towards the outer regions while \nnhs traces the denser inner parts. 

\subsubsection{Drop profile}
To test the effect of CO chemistry we use a `drop' profile as described in \cite{Jorg04}. In its simplest implementation, the model assumes a constant, undepleted abundance of CO relative to that of H$_2$, X$_0$, where $T_{\textrm{dust/gas}} \geq T_{\textrm{evap}}$ or $n$(H$_2$) $< 3 \times 10^4$ cm$^{-3}$ combined with a depleted abundance of X$_D$, elsewhere. Using equation (\ref{eqn:Tgas}), it is possible to reformulate the model solely in terms of the local gas density, $\rho$,
\begin{equation}
X(\rho) = \left\{
\begin{array}{l l}
    X_0 & \textrm{for } 10^{-19} <\rho< 1.56 \times 10^{-12} \textrm{ g cm}^{-3},\\[1.25ex]
    X_\textrm{D} & \rm{elsewhere},
\end{array}
\right.
\label{eqn:dropprofile}
\end{equation}
Empirically, the drop profile has been used to improve the fit of  Gaussian line profiles to low $J$ observational \highs data of various class 0, class 1 and pre-stellar objects. The improvement in the goodness of fit is found to be sufficiently significant as to provide strong evidence for its acceptance over the constant abundance model. Moreover, the model has its basis in observations \citep{Caselli99, Tafalla02} where a freeze-out zone has been directly imaged.

Physically, the profile is explained in terms of radii. The material within the inner radius is expected to be gas-phase owing to its proximity to the warm core. The canonical temperature at which CO is thought to have returned to the gas-phase is 30\,K. Beyond the outer radius, at densities lower than $3 \times 10^4$ cm$^{-3}$, the depletion timescale due to freezeout is thought to exceed the lifetime of the core and is thus assumed to remain at an undepleted abundance, to reduce the number of free parameters. 

Since the critical densities of \nnhs are higher than those of the CO isotopologues, they are less sensitive to the outer region of the envelope where depletion and contribution from the surrounding cloud may be important. It has been posited (\citealt{Jorg04}) that this may explain why these transitions can be modelled assuming a constant density. 

In our chemistry runs we repeat the same method as outlined in section \ref{discretisation} but with depleted densities one order of magnitude lower in the intermediate region, viz. [\low]$_\textrm{D}$~=~$1~\times~10^{-7}$ and [\high]$_\textrm{D}$~=~$1~\times~10^{-8}$. 
\subsection{Analysis}
\label{analysis}

For each synthetic observation the intensity datacubes had their background intensity subtracted, $B(\nu_0,T_{\textrm{cmb}})$, and were summed over all velocity channels to give an integrated intensity map. The maps were converted to brightness temperature, T$_{\textrm{b}}$ using the Rayleigh-Jeans approximation, i.e. 
\begin{equation}
\label{eqn:Tb}
T_{b} = \frac{c^2}{2k\nu^2} I_{\nu}.
\end{equation}

We used \textsc{wavdetect}, part of the \textsc{ciao} data analysis system written for the Chandra X-ray Observatory, to detect cores in the \nnhs intensity maps. The \textsc{wavdetect} tool \citep{FreKasRos02} correlates the image with wavelets of different, user-defined scales and then searches the results for significant correlations. This method is very similar to the difference of Gaussians (DoG) method, where a difference map is created by convolving an image with two Gaussian distributions with different spatial variances and subtracting the resultants, preserving structure with lengthscales between the two variances. 
\textsc{wavdetect} produced very similar results but was slightly better at detecting objects that it was not possible to resolve using DoG and was more robust against the converse case of rejecting objects that upon later inspection had distinct line profiles (often indicative of containing protostars). We looked for cores with lengthscales of 2 -- 64 pixels ($\sim 0.001-0.04$ pc) and applied a threshold of 50 per cent of the global maximum to the transformed image to determine the membership of a pixel to a core. A recursive blob counting algorithm was invoked to determine the number of separate regions identified as containing a local maximum within the map; for \nnh, each region is a potential core and for \lows and \highs each region defines the envelope which surrounds the core. This is similar to the method of \cite*{Offneretal08} in that we generate a background map and then detect local intensity peaks above that. 

Any core candidate that lay wholly or partially outside the enveloping material as determined by the low density contour was excluded from the analysis. In this constant abundance analysis, it is expected that the regions should be strict subsets of each other. For each core candidate, a mask was applied to the original datacube corresponding to the region defined by the 50 per cent contour and line profiles for that region were found by integrating over all pixels in the region for each molecular tracer. 

\begin{figure*}
\centering
\includegraphics[width=1.0\textwidth]{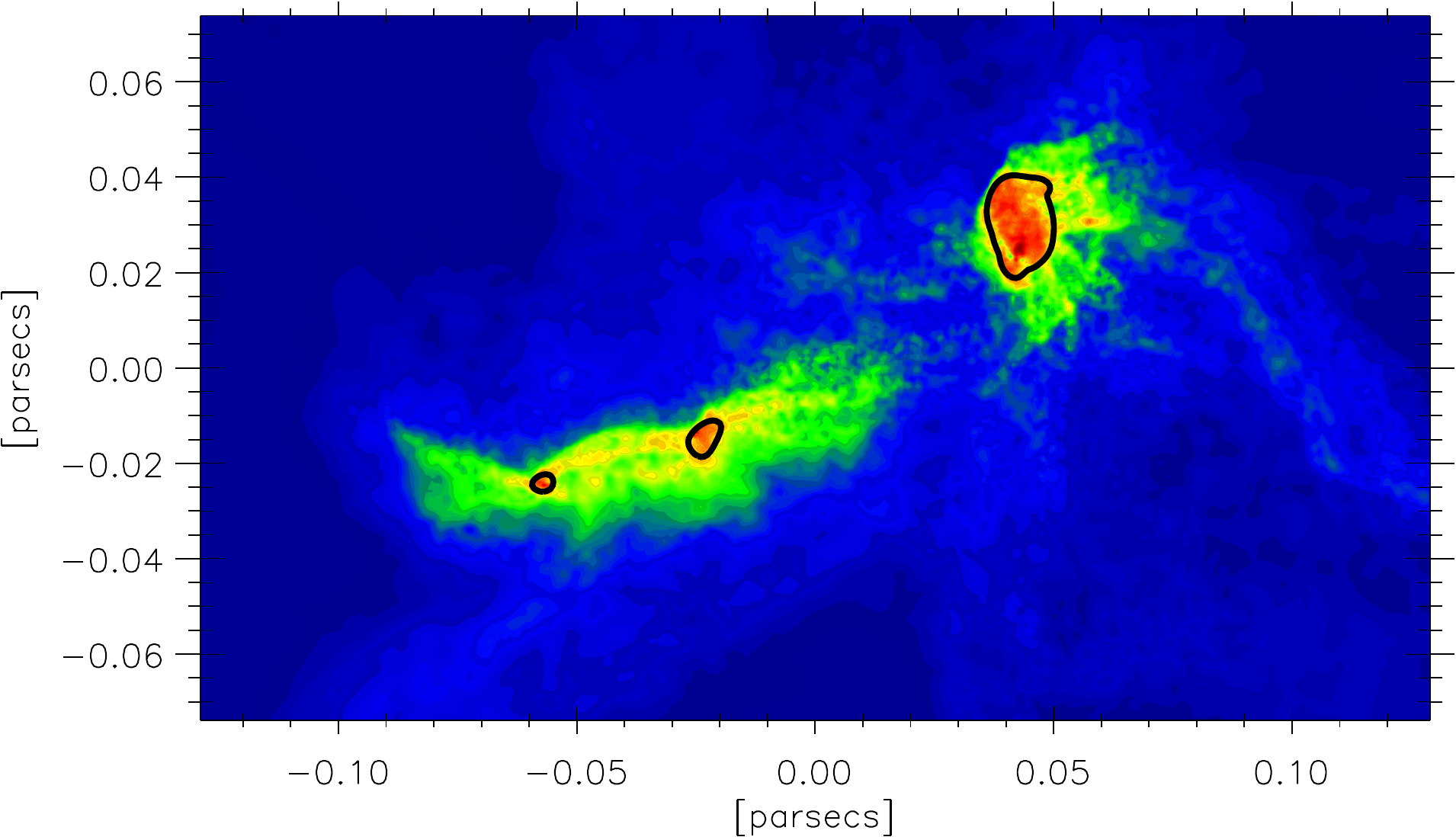}
\caption{Integrated intensity map in \nnh~(1-0). Contours have been added to demarcate each of the cores identified. The line profiles of the two largest cores and those of the enveloping gas are illustrated in Figures \ref{fig:coreprofiles} and \ref{fig:envelopeprofiles} respectively. Note that this image only shows the region of interest in the cloud. The extent of the entire cloud is $\approx 0.6$ pc.}
\label{fig:map}
\end{figure*}

\begin{figure*}
\centering
\subfigure{
\includegraphics[width=0.4\textwidth]{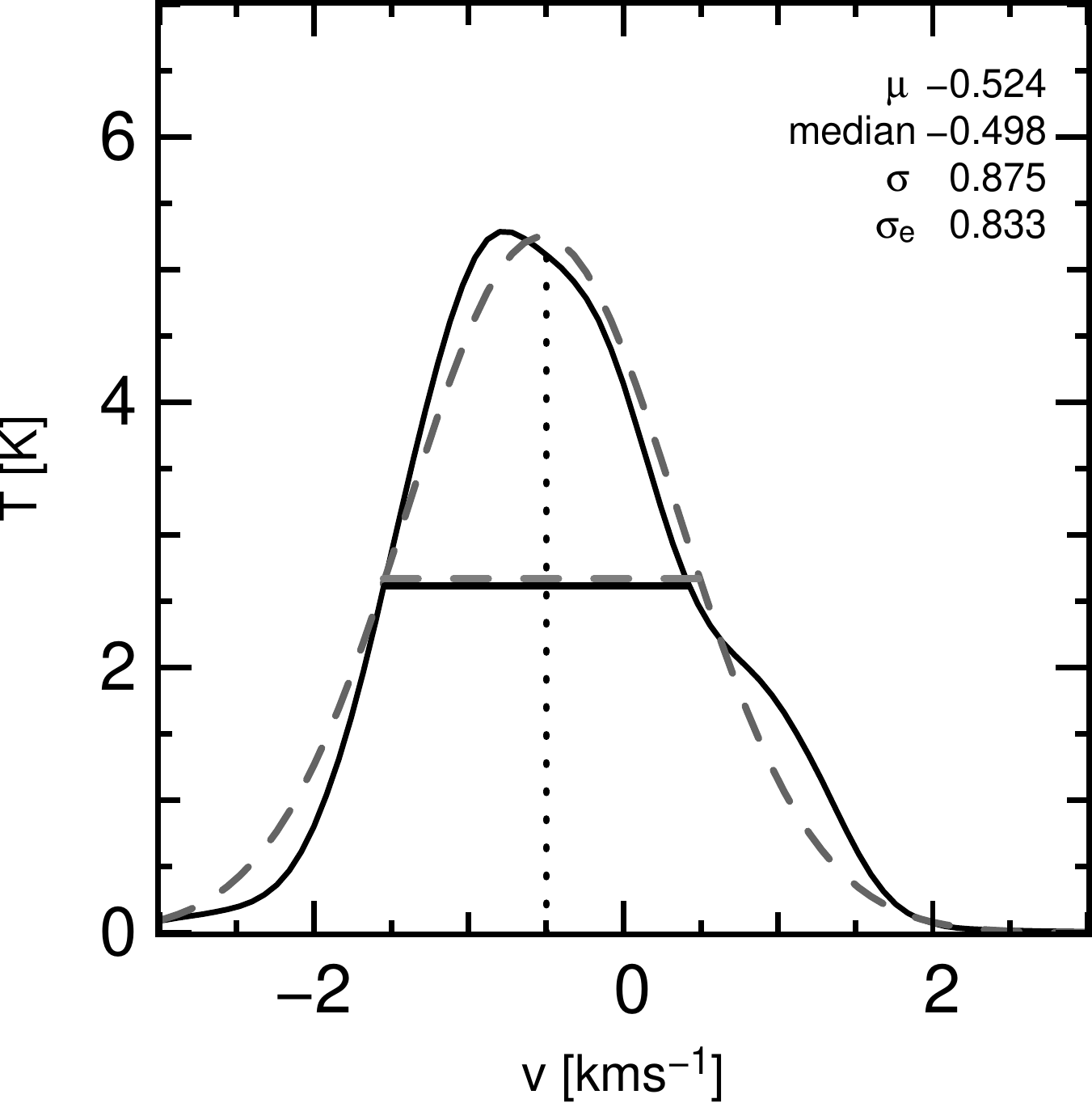}
\label{fig:subfig1}
}
\subfigure{
\includegraphics[width=0.4\textwidth]{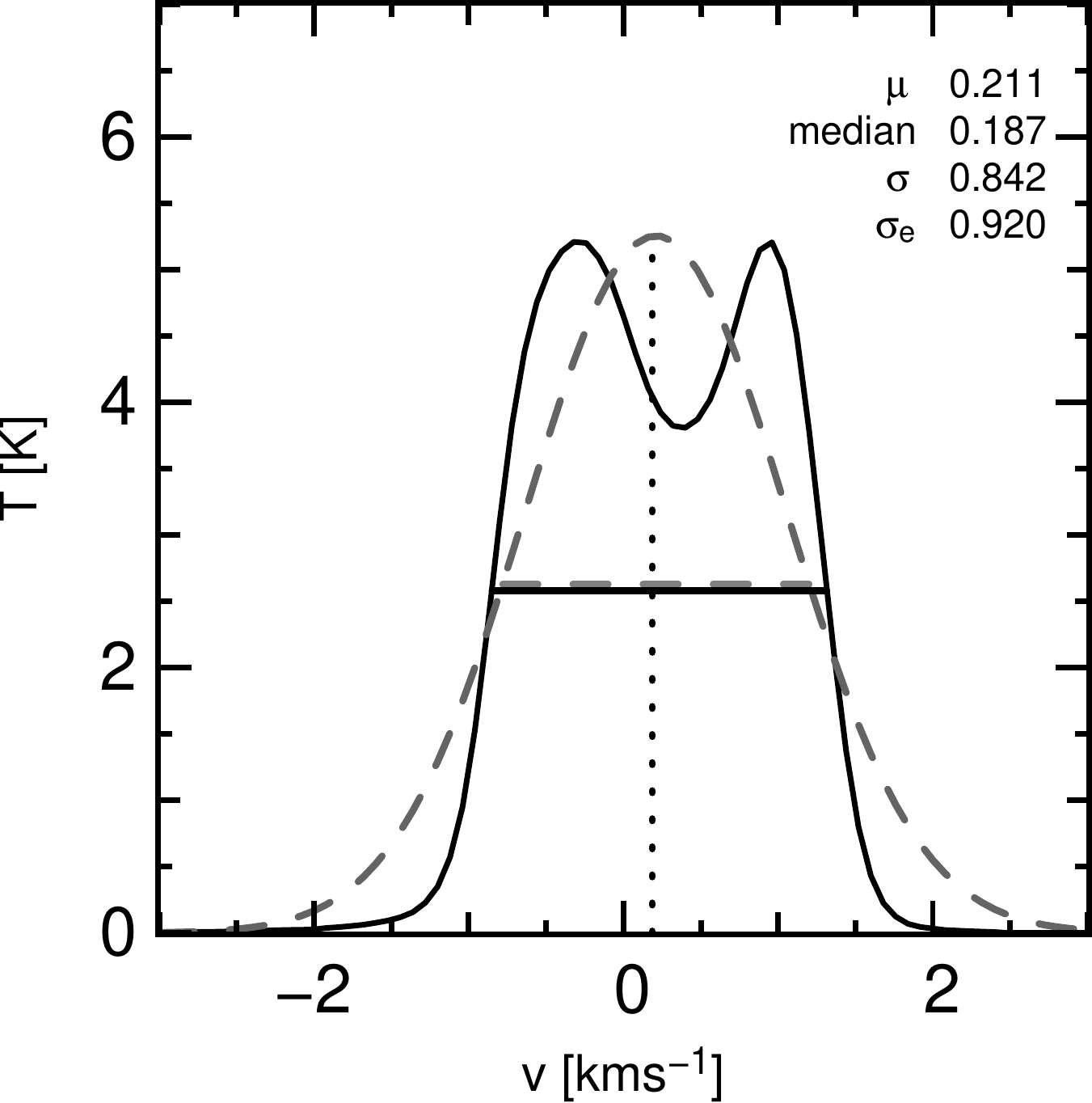}
\label{fig:subfig2}
}
\caption{Typical core profiles. Dashed lines indicate the Gaussian profile fit and FWHM. Dotted line indicates position of the Gaussian line centre. The legend also displays median line centre and the equivalent standard deviation, $\sigma_{e}$, from the FWHM of the profile. \subref{fig:subfig2} shows a profile that it would be possible to fit two Gaussians to (assuming two cores along the line-of-sight) however by examining the SPH particle distribution this profile is known to be caused by rotating gas surrounding a multiple system.}
\label{fig:coreprofiles}
\end{figure*}

\begin{figure*}
\centering
\subfigure{
\includegraphics[width=0.4\textwidth]{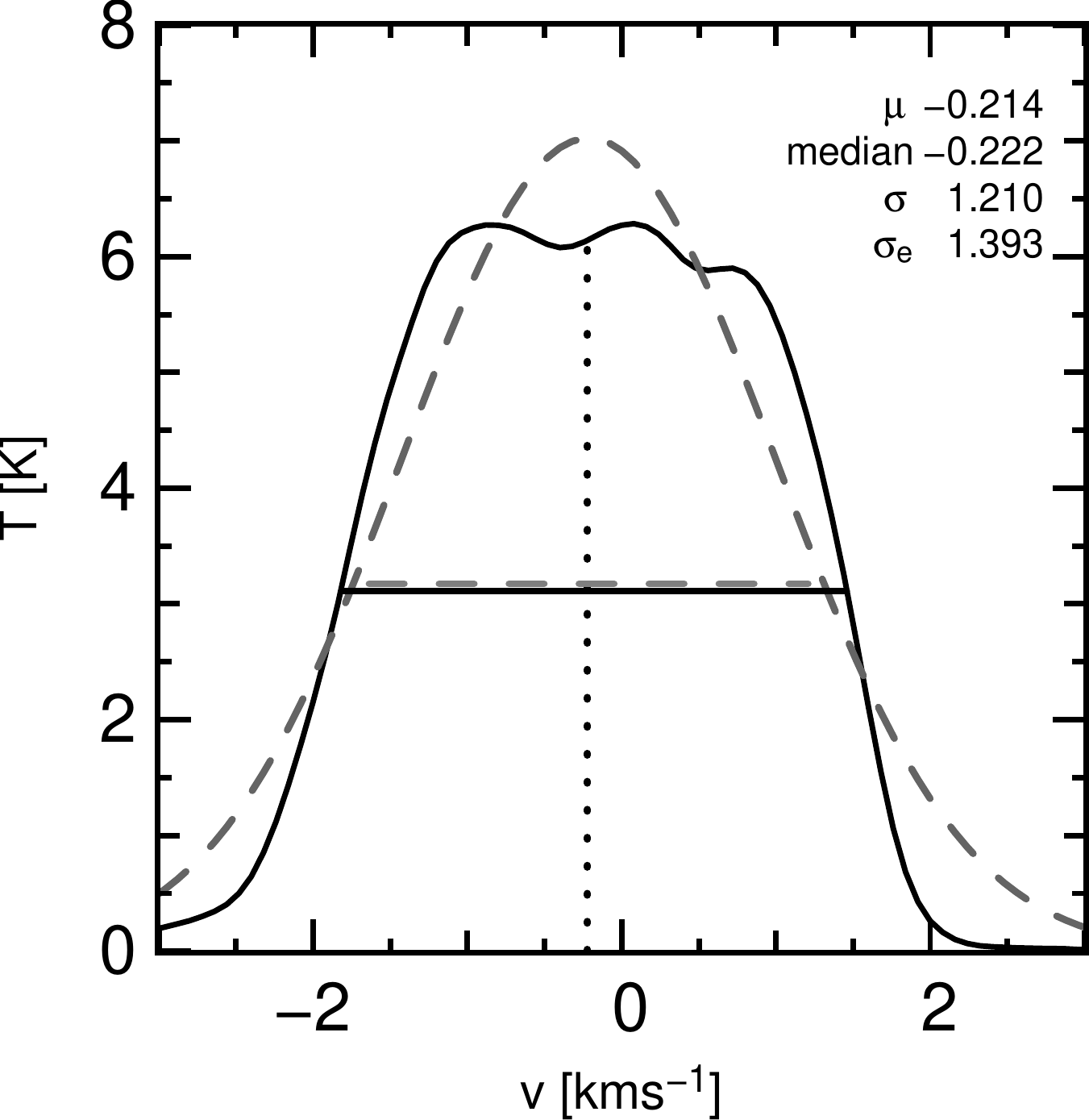}
\label{fig:subfig1}
}
\subfigure{
\includegraphics[width=0.4\textwidth]{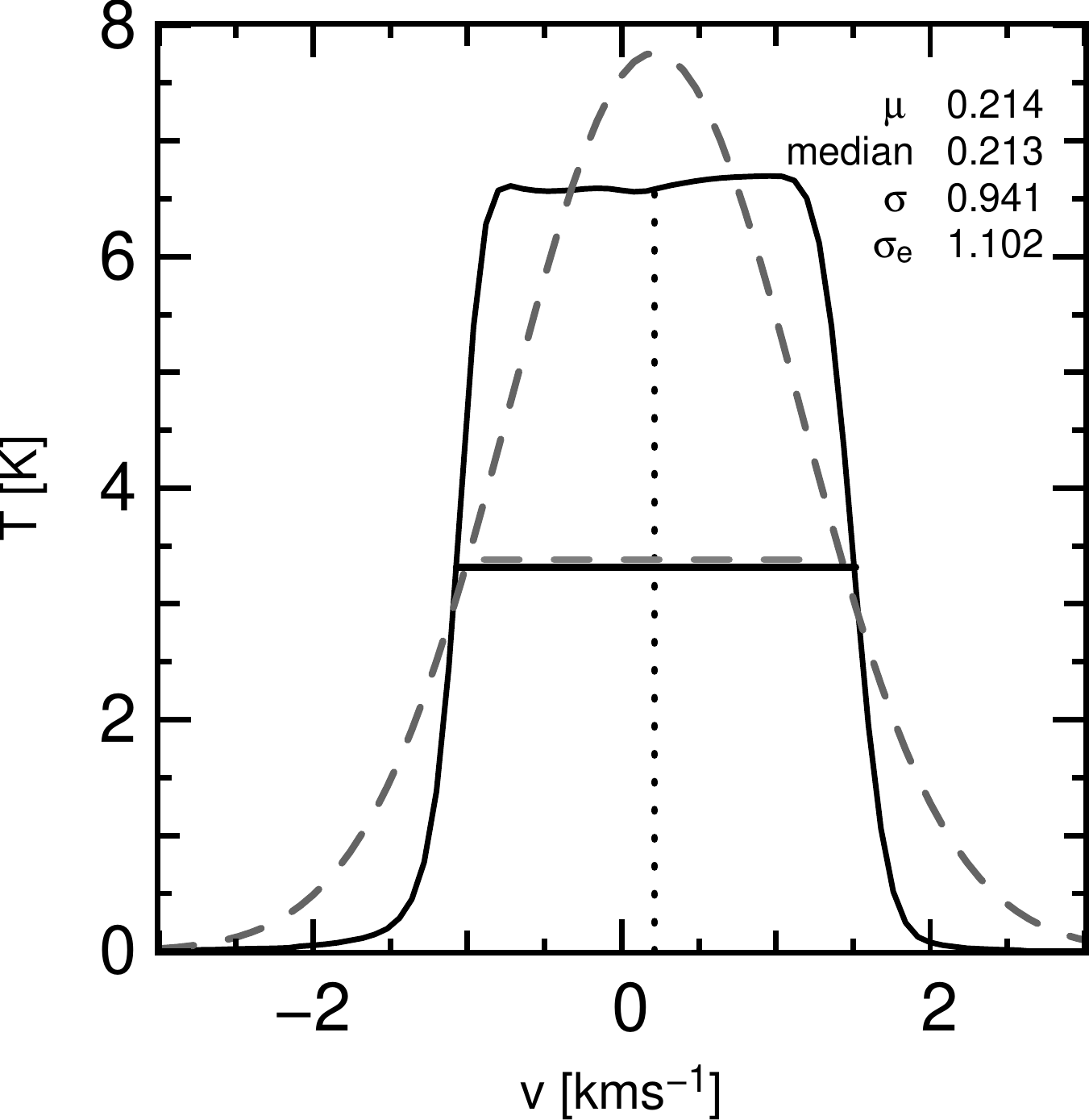}
\label{fig:subfig2}
}
\caption{Typical low-density envelope profiles traced using \low. Both profiles exhibit flat peaks associated with the saturation of optically thick \low~(1-0) line.}
\label{fig:envelopeprofiles}
\end{figure*}

The line profile was fitted with a Gaussian distribution using a Levenberg-Marquardt algorithm. The initial Gaussian was set to have a peak brightness temperature of 5~K, $\sigma = 1$ km s$^{-1}$ and to be centred at $v = 0$ km s$^{-1}$. Each datum point was assumed to have equal weighting. In addition to characterizing the line profile using those parameters, the mean and median of the distribution were calculated providing a robust method to verify the position of the peak in the Gaussian distribution determined by fitting. Similarly, the full-width at half-maximum (FWHM) of the profile was recorded to parameterize the variance. The same process was repeated for the low density tracer and the resultant difference in line centre velocities was assumed to represent the systematic shift in line-of-sight velocity. Figures \ref{fig:map}, \ref{fig:coreprofiles} and \ref{fig:envelopeprofiles} illustrate the results of the analysis for one particular observation at 1.4 t$_{\textrm{ff}}$.

\subsection{Results}
\label{results}

\begin{figure*}
\includegraphics[width = 0.78 \textwidth]{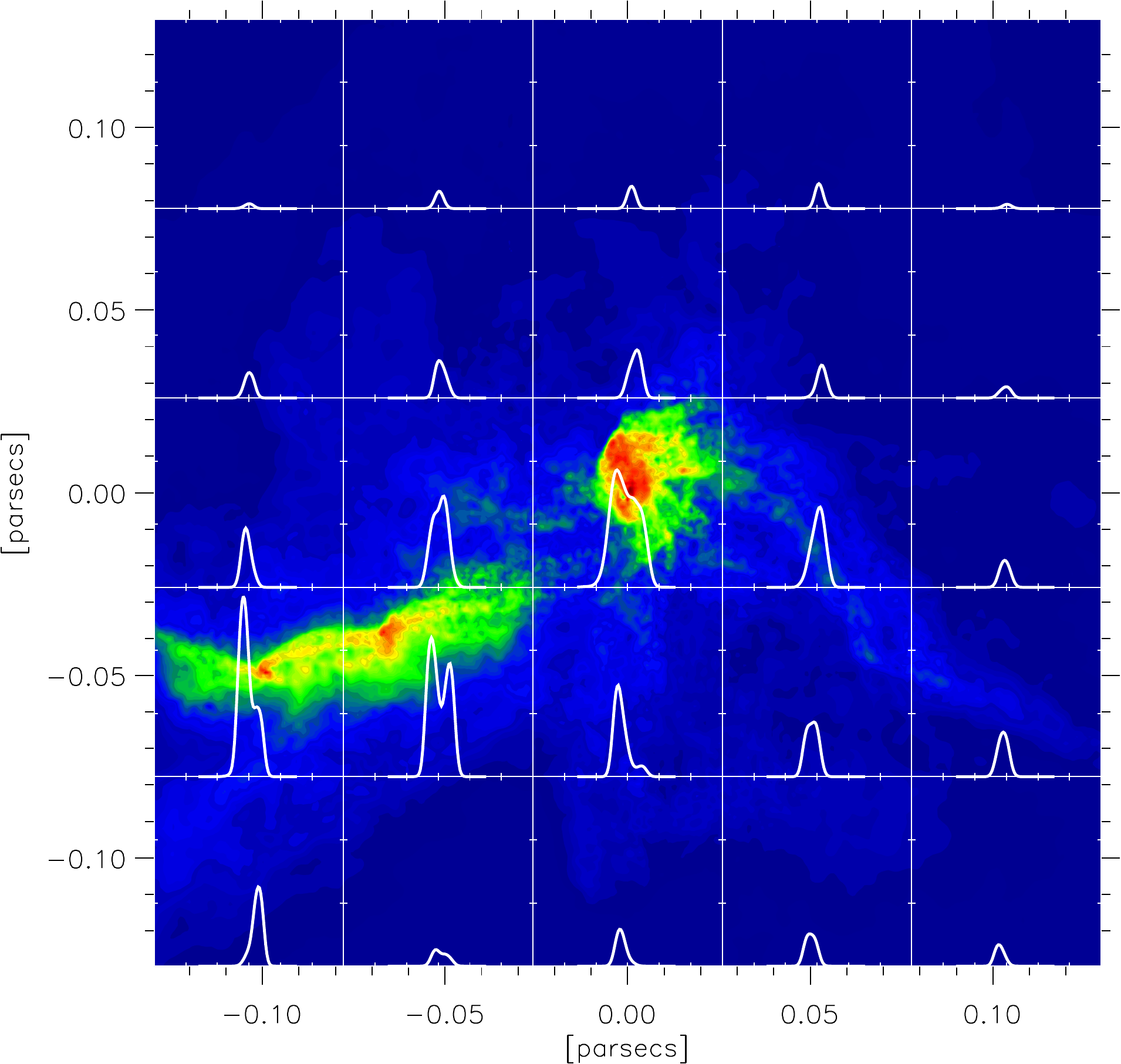}
  \caption{The line profiles of the square regions are superimposed onto a \nnh~(1-0) integrated intensity map (see Figure \ref{fig:map}). The regions containing the three cores have somewhat broader line profiles and exhibit more structure (i.e. double peaks caused by rotation and infall) than the optically thin gas surrounding them.}
  \label{fig:5x5map}
\end{figure*}

Having removed the two cores that did not lie within the 50 per cent low-density envelope we also removed 11 cores (over the 5 timesteps) which were judged to be spurious detections which were typically characterized by very small areas ($<10$ pixels), similar velocity profiles to a larger nearby detection and often did not contain any sink particles. This left 169 cores over the 5 timesteps; the fewest detected cores (22 from 20 observing positions) being in the first timestep before the creation of any sink particles and the most (58) being detected in the last timestep, where three unique star-forming regions were detected by \cite{BaBoBromm03}. Every remaining valid core-envelope pair was analyzed and the results were collated to allow us to perform a statistical analysis according to the strategy outlined in section \ref{analysis}. Cores that contained one or more sink particles were judged to be protostellar cores while cores without sink particles are categorized as being prestellar (or starless).
Following \cite{Ayliffeetal07}, we do not present the results of Gaussian fits in this work. All analysis was done using both methods and the differences between methods of analysis were sufficiently small that they did not affect any conclusions drawn upon the results of the work. Moreover, the profiles of \lows data were rarely found to be Gaussian, exhibiting flat, wide peaks (owing to their large optical depth). As such, we characterize the dispersion using the FWHM / $2\sqrt{2\textrm{ln}(2)}$ of any given line profile and take the line centre to be the median, $m$, of the distribution ($\int_{-\infty}^{m} f(x)$ d$x$ = 0.5 $\int_{-\infty}^{\infty} f(x)$ d$x$). As a non-parametric estimator of the line centre, the median does not assume normally distributed data; as a result the median is less influenced by shoulders, long tails, noise etc. and may give a less biased estimate of the true line centre. Consequently we did not attempt to fit multiple Gaussians to the data (with the assumption that a double-peaked distribution indicated two cores along the line-of-sight) primarily because it was possible to confirm from the SPH particle distribution that this was not the case but also because it allows us to more directly compare our results with previous studies. The presence of a doubly (or more) peaked profiles is caused by the velocity shift associated with the rotation and/or infall of gas. In order to illustrate these effects we have calculated profiles from 25 areal bins around the cores shown in Figure~\ref{fig:map} and plotted them over an integrated intensity map (Figure~\ref{fig:5x5map}). Whilst the optically thin gas is associated with single-peaked profiles, the denser cores display a more complex line profile morphology. A similar variety of profile shapes have been found in one-dimensional simulations of collapsing and rotating cores \citep{Pavetal08}.

\begin{figure}
\includegraphics[width = 0.475 \textwidth]{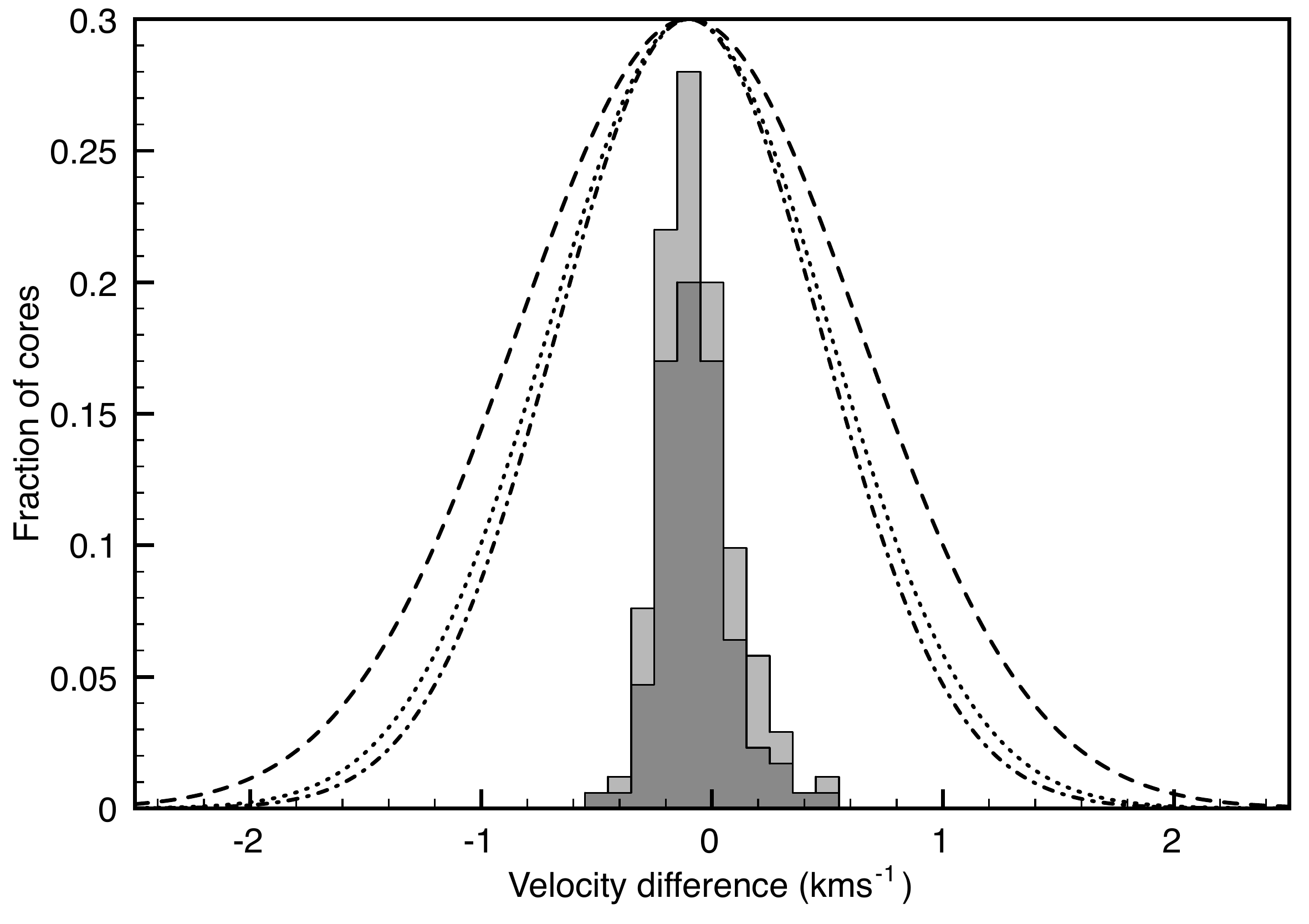}
  \caption{Histogram of velocity difference of all cores (combined starless and protostellar cores) identified at all time steps (light grey, $\sigma = 0.16$ km s$^{-1}$) and protostellar cores (dark grey, $\sigma = 0.18$ km s$^{-1}$). Also plotted are the velocity dispersions for \nnhs (dot, $\sigma = 0.861$ km s$^{-1}$), \highs (dot-dash, $\sigma = 0.808$ km s$^{-1}$), \lows (dash, $\sigma = 1.05$ km s$^{-1}$).}
  \label{fig:allcores}
\end{figure}
Figure~\ref{fig:allcores} shows a histogram of the distribution of relative velocity difference between \nnhs cores and low density gas envelopes identified by \lows summed over all timesteps. The standard deviation of all cores (both pre-- and protostellar) is $\sigma=0.16$ km s$^{-1}$. 
In the same figure, we plot three normalized Gaussian distributions whose standard deviations are equal to that of the equivalent standard deviations as derived from the mean FWHM of the line profiles for each tracer molecule. They are notably more broad than either the relative velocity difference or the sound speed.

\begin{figure}
\includegraphics[width = 0.475 \textwidth]{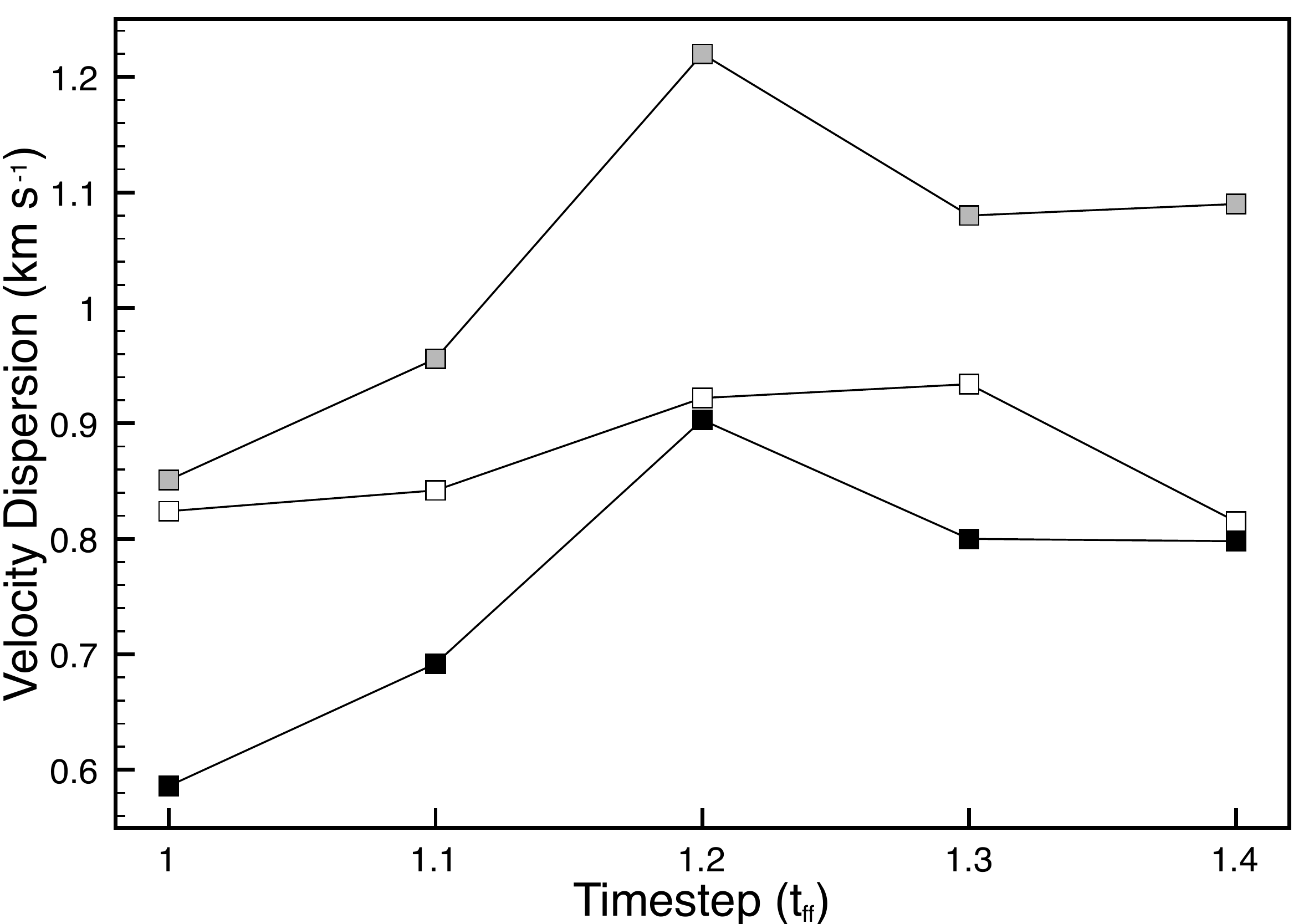}
  \caption{The velocity dispersion of each molecular species for each timestep. White squares denote \nnh, grey squares denote \lows and black squares denote \high.}
  \label{fig:timedep}
\end{figure}

Figure~\ref{fig:timedep} illustrates values for the mean velocity dispersions of each tracer for each timestep. The trend for the CO isotopologues is the same; that the mean velocity dispersion is relatively low at 1.0 t$_{\textrm{ff}}$, rises until a peak at 1.2 t$_{\textrm{ff}}$ and then turns over and levels off towards the end of the simulation. The same effect can be seen in the \nnhs cores except that the effect seems to be delayed until 1.3 t$_{\textrm{ff}}$. We discuss this trend in section \ref{discussion}.
The relative differences in line-centre velocities between the \highs envelope and the \nnhs cores are lower throughout the cluster. This is due to \highs emission being more optically thin than the \lows emission. Consequently, \high~(1-0) is able to trace a region closer to the core. Note that in these constant abundance calculations CO is not frozen-out and therefore will trace a higher density region. 

\subsubsection{The effect of chemistry}
We repeated the analysis with chemical model outlined in section \ref{chemistry}. Although minor differences in line profile intensity and shape are observed, the velocity dispersions did not change significantly and no systematic differences with the constant abundance models were detected. It appears that the optical depths of the transitions we have studied are such that the effective photo-surface of the cores is outside the volume where freeze-out occurs. Naturally a more sophisticated treatment of chemistry, involving a full time-dependent calculation, is required to confirm our preliminary results. However such a calculation is well beyond the scope of this paper.

\section{Discussion}
\label{discussion}

We performed a detailed radiative transfer analysis of a self-gravitating hydrodynamics simulation in order to examine the relationship between the core and envelope velocity dispersions and make a more direct comparison between the hydrodynamical models and millimetre observations. We used information about specific molecular tracers and determined the non-LTE level populations and used these to calculate the emergent intensity from different observation angles around the cluster. We found good qualitative agreement with observational results (e.g. \citealt{Walshetal04,Walshetal07,Kirketal07}), our main conclusion being that one cannot reject competitive accretion as a viable theory of star formation based on observed velocity profiles.
\cite{Ayliffeetal07} found that the majority of their sources had velocity differences (between high and low density gas) less than the sound speed at all timesteps but that they had a significant tail out to large velocity differences (in this case $> 0.5$ km~s$^{-1}$). However, they also found that their high-density gas linewidth became larger than their low-density gas linewidth at times greater than 1.1~t$_{\textrm{ff}}$, contradicting the observations of \citeauthor{Walshetal04} and \cite*{Kirketal07}. We believe this may be because the study done by \cite{Ayliffeetal07} did not take into account optical depth effects and that their density cut for the high density material was too high. By reducing their assumed critical density by a factor of 3 they were able to reduce the velocity dispersion of high-density material by a significant amount so that the high-density gas linewidth was smaller than the low-density linewidth, when averaged over all timesteps. This work does not show the trend of larger high-density linewidths at any time, although Figure~\ref{fig:timedep} shows that the velocity dispersion turns over with increasing time. We believe that this effect can be explained by two opposing processes. The original simulation did not drive turbulence so it is expected that in the absence of dynamic interactions between protostellar objects the velocity dispersion of the gas will ultimately tend to decrease. However, as we are observing regions undergoing complex interactions that act to stir up the gas, thereby increasing the velocity dispersion, we see the two effects combined; each star formation episode acting to increase the velocity dispersion and decay over time acting to reduce it. Like \citeauthor{Ayliffeetal07} and the observational surveys, we see no clear trend in the relative motions of core/envelope pairs and note that the standard deviations of the velocity differences are always far smaller than the velocity dispersion.

The original hydrodynamic simulation did not include any form of driven turbulence. We have assumed a global, constant, non-turbulent line width of 0.3 km s$^{-1}$. This is similar to that of the mean non-thermal turbulent line width of the driven case in \citet{Offneretal08b} but at odds with the theory that cores are supported by thermal pressure because large non-thermal motions will have disappeared on small scales ($< 0.1$ pc) and at high densities ($> 10^4$ cm$^{-3}$) \citep{JohRosTaf10a}. As a result we found larger velocity dispersions than have been observed in some studies \citep{Offneretal08}.  However, not only are our observed core motions smaller than the mean velocity dispersion but also they are often less than the local sound speed (Figure~\ref{fig:allcores}). This is true for both starless and protostellar cores and in fact, our results mirror those of \citeauthor{Kirketal07} in that they too show no indication that the core-to-envelope motions significantly change between the starless and protostellar stages of evolution. By performing a more detailed radiative transfer study of \citeauthor{BaBoBromm03}'s cluster simulation, our findings go some way towards reconciling the results previously obtained by \cite{Ayliffeetal07} with those found by others who had performed similar studies. 

Although relatively sophisticated we intend to incorporate additional physics in future. Specifically we have not yet taken into account \nnhs hyperfine splitting. \cite{Tafallaetal04} studied the effect of neglecting hyperfine splitting (which maximizes radiative trapping). They found that the difference in obtained level population was on the order of tens of per cent which is comparable to the uncertainties in collision parameters. Neglecting hyperfine splitting can have a significant effect on the optical depth of any \nnhs transition and consequently the emergent intensity. In effect, we have systematically overestimated the line optical depth and the inclusion of hyperfine splitting would shift the line formation regions to higher density. In the future we wish to incorporate hyperfine splitting, although previous work \citep{DanCerDub06} incorporating this microphysics has demonstrated that whilst it is likely to have a significant effect it will not be large enough to affect our overall conclusions.

Another important factor is depletion of CO by adsorption onto dust grain surfaces. Throughout this paper, we have assumed that all tracer molecules occur with constant abundance and many observational studies have shown that this is demonstrably not the case. Chemical networks \citep{Bergin97,GloMac07a} exist that allow us to predict the abundance of many molecular species in molecular clouds. We used a simple chemical model to estimate the effect that CO chemistry would have on our observations, and have discovered that a simple freeze-out model does not significantly affect our results. Nonetheless we recognize that a more complete treatment is desirable and a natural extension of this work would be to couple \torus with a chemical network to understand more fully where line emission comes from in molecular clouds. 
\section{Conclusion}
\label{conclusion}
This paper demonstrates the molecular line radiative transfer capabilities of \textsc{torus}, a radiative transfer code that uses AMR to resolve the fine details of complex three-dimensional environments. The molecular line transfer module was compared against other codes in a scenario where LTE conditions did not apply and the results showed excellent agreement. 

In environments such as molecular clouds, gas that is dense enough to emit with sufficient intensity in a given molecular line may only exist in small regions when compared with the size of the whole cloud. Consequently, fine discretization is only required over a small volume. Thus, we devised an efficient method of mapping irregular SPH data onto an AMR grid with minimal discretization error. This technique permitted the study of the expected molecular line emission from a hydrodynamical simulation of molecular cloud collapse using SPH with our grid-based radiative transfer code. Having obtained non-LTE level populations throughout the cluster for tracers of high-- and low--density gas over 5 timesteps covering the creation of 50 protostellar objects we were able to produce and analyse synthetic observations. We have shown in this paper that clusters exhibiting competitive accretion are able to reproduce the properties of relative core motions found in observation. 

As well as enabling us to accurately model line profiles, the mapping of density distributions from the particle-based representation onto an AMR grid has important implications for coupling radiation transfer and hydrodynamics. For example, using the technique outlined in section \ref{sphtogrid}, it is now possible to conduct a full, polychromatic multiple scattering treatment of the transport of radiation in an SPH simulated circumstellar disc using high-accuracy grid-based techniques \citep{AcrHarRun10} without having to resort to the flux-limited diffusion approximation employed in earlier radiative transfer simulations (e.g. \citealt{WhiteBate04,KruKleMcK07a,Bate09b}). However, extending such calculations to a cluster collapse model is currently not tractable given the very large dynamic range of linear scales necessary to accurately treat the radiation transport of optically thin-to-thick boundaries, but, with further advances in computational power this kind of calculation will become a reality, bringing with it enhanced certainty in the conclusions that can be reached using these simulations.

\section{ACKNOWLEDGMENTS}

DAR would like to thank Ben Ayliffe, Nathan Mayne and Eric Saunders for useful discussions and helpful comments on this work. The computations by \citeauthor{BaBoBromm03} were performed using the U.K. Astrophysical Fluids Facility (UKAFF) and the radiative transfer analysis of the calculation was performed using the University of Exeter supercomputer, an SGI Altix ICE 8200. DAR was supported by an STFC-funded post-graduate studentship. We thank \citeauthor{Schoier05} for making the LAMDA database publicly available and the anonymous referee for their detailed and insightful comments on our original manuscript.

\bibliographystyle{mn2e}
\bibliography{paper}
\appendix
\label{lastpage}
\end{document}